\documentclass[fleqn,usenatbib]{mnras}

\usepackage{newtxtext,newtxmath}

\usepackage[T1]{fontenc}

\usepackage{xcolor}
\usepackage{soul}

\usepackage{graphicx}	
\usepackage{amsmath}	
\usepackage{anyfontsize} 

\definecolor{bluehl}{rgb}{0.75,0.75,1}

\title[The colors of TNG100]{The many colors of the TNG100 simulation}

\author[A. Gebek et al.]{%
Andrea Gebek,$^1$\thanks{E-mail: andrea.gebek@ugent.be}
Ana Trčka,$^1$
Maarten Baes,$^1$
Marco Martorano,$^1$
Annalisa Pillepich,$^2$
\newauthor
Anand Utsav Kapoor, $^{1,3}$
Angelos Nersesian, $^{1,4}$
and Arjen van der Wel $^1$
\\
$^1$Sterrenkundig Observatorium, Universiteit Gent, Krijgslaan 281 S9, 9000 Gent, Belgium\\
$^2$Max-Planck-Institut für Astronomie, Königstuhl 17, D-69117 Heidelberg, Germany\\
$^3$ Laboratoire Lagrange, Université Côte d’Azur, Observatoire de la Côte d’Azur, CNRS, Blvd de l’Observatoire, CS 34229, 06304 Nice cedex 4, France\\
$^4$ STAR Institute, Université de Liège, Quartier Agora, Allée du Six Août 19c, 4000 Liège, Belgium
}

\date{Accepted XXX. Received YYY; in original form ZZZ}

\pubyear{2024}

\begin{document}
\label{firstpage}
\pagerange{\pageref{firstpage}--\pageref{lastpage}}
\maketitle

\begin{abstract}
We apply the 3D dust radiative transfer code SKIRT to the low-redshift ($z\leq0.1$) galaxy population in the TNG100 cosmological simulation, the fiducial run of the IllustrisTNG project. We compute global fluxes and spectral energy distributions (SEDs) from the far-ultraviolet to the sub-millimeter for $\approx60\,000$ galaxies. Our post-processing methodology follows the study of Trčka et al. (2022) of the higher-resolution TNG50 simulation. We verify that TNG100 reproduces observational luminosity functions at low redshifts to excellent precision, unlike TNG50. Additionally, we test the realism of our TNG100 plus SKIRT fluxes by comparing various flux and color relations to data from the GAMA survey. TNG100 broadly reproduces the observed distributions, but we predict ultraviolet colors that are too blue by $\approx0.4\,\mathrm{mag}$, possibly related to the extinction in the star-forming regions subgrid model not being selective enough. Furthermore, we find that the simulated galaxies exhibit mid-infrared fluxes elevated by up to $\approx0.5\,\mathrm{mag}$ that we attribute to overly effective stochastic heating of the diffuse dust. All synthetic broadband fluxes and SEDs are made publicly available in three orientations and four apertures, and can readily be used to study TNG100 galaxies in a mock observational fashion.

\end{abstract}

\begin{keywords}
methods: numerical -- galaxies: photometry -- galaxies: evolution -- ISM: dust, extinction -- radiative transfer
\end{keywords}



\section{Introduction}

Cosmological hydrodynamical simulations that emulate the assembly and evolution of thousands of galaxies have proven an indispensable tool to understand many facets of the observed galaxy population (\citealt{Somerville2015}; \citealt{Vogelsberger2020}). Assessing the realism and reliability of cosmological simulations by comparing their outcome to observations is critical: a solid baseline agreement is necessary in order to draw meaningful conclusions from the simulations. Furthermore, discrepancies can be used to unveil gaps in our understanding of galaxy formation and evolution. However, comparing the simulated and observed galaxy populations comes with a major caveat: observations of galaxies only trace the light emitted by stellar populations (and partially reprocessed by dust and gas), while in simulations only the `physical' parameters of the stellar populations and interstellar medium (such as masses and metallicities) are known. Tracking the radiation field in cosmological simulations is computationally prohibitive, unless the simulation run only covers the high-redshift regime ($z\gtrsim5$) and only few wavelength bins are considered as is done in the SPHINX (\citealt{Rosdahl2018}) or THESAN (\citealt{Kannan2022}) simulations. 

Comparing the simulated and observed galaxy populations in the `physical' realm (e.g. the stellar mass function or the main sequence of star-forming galaxies) bears the main caveat that all physical properties need to be inferred from observations. Such retrievals of physical parameters sensitively depend on the adopted model used in e.g. the SED fitting process, relying on simplified star-formation histories and dust-to-star geometries (\citealt{Pacifici2023}). As an example cautionary note, the long-standing disagreement in the star-forming main sequence for $0.5<z<3$ with the simulated galaxy population offset to lower star-formation rates (\citealt{Mitchell2014}; \citealt{Leja2015}; \citealt{Furlong2015}; \citealt{Tomczak2016}; \citealt{Donnari2019}; \citealt{Katsianis2020}) could only recently be remedied with more sophisticated SED fitting methods (\citealt{Nelson2021}; \citealt{Leja2022}). 

As a complementary approach, it is therefore critical to move the simulated galaxies into the observational realm by postprocessing them with radiative transfer. This method circumvents any uncertainties in the parameter inference from observations (e.g. choice of free parameters and prior ranges), but requires a postprocessing scheme based on the stars and gas of the simulated galaxies that comes with its own caveats (e.g. choice of dust allocation recipe if dust is not modelled in the cosmological simulation). As a substantial fraction of the light emitted by stellar populations is reprocessed by dust and gas in the interstellar medium (\citealt{Popescu2002}; \citealt{Viaene2016}; \citealt{Bianchi2018}), methods to solve for the transport of radiation are required. Since dust efficiently scatters and absorbs starlight at ultraviolet (UV) and optical wavelengths, Monte Carlo radiative transfer (MCRT) methods are generally used to accurately simulate the radiation field in galaxies taking the 3D dust and stellar distributions into account. Using such MCRT methods, synthetic broadband fluxes and images for a large variety of cosmological simulations such as EAGLE (\citealt{Camps2016}; \citealt{Trayford2017}), SIMBA (\citealt{Narayanan2021});  AURIGA (\citealt{Kapoor2021}; Kapoor et al. in prep.); ARTEMIS (\citealt{Camps2022}), IllustrisTNG (\citealt{Rodriguez-Gomez2019}; \citealt{Schulz2020}; \citealt{Vogelsberger2020b}; \citealt{Trcka2022}; \citealt{Popping2022}; \citealt{Costantin2023}; \citealt{Guzman-Ortega2023}; \citealt{Baes2024}; \citealt{Bottrell2024}), and NewHorizon (\citealt{Jang2023}) have been calculated and compared to observational data.

In \citet{Trcka2022}, broadband fluxes from the UV to the far-infrared (FIR) have been computed for a sample of $\sim$ 14\,000 galaxies at low redshift ($z\leq0.1$) from the TNG50 simulation. Comparing the simulated fluxes to observational low-redshift luminosity functions (LFs), \citet{Trcka2022} found that the TNG50 LFs at all wavelengths exceed the observational estimates. \citet{Trcka2022} attribute 
this tension mostly to the subgrid parameter calibration in the IllustrisTNG project. As some of the physical processes cannot be resolved by cosmological simulations, these simulations typically rely on a number of subgrid parameters (e.g. the strength of feedback from active galactic nuclei) to reproduce some important galaxy statistics (e.g. the stellar mass-halo mass relation, \citealt{Schaye2015}; \citealt{Kugel2023}). In the case of the IllustrisTNG project, these subgrid parameters were chosen at the resolution of the fiducial TNG100 run and then left constant for other simulation runs. This leads to small systematic resolution-dependent differences in the outcomes of the various IllustrisTNG runs (see Appendices of \citealp{Pillepich2018, TNG_Pillepich, Pillepich2019} for more details).

In this study, we want to test if the IllustrisTNG subgrid choices truly caused the discrepancies between TNG50 and observations found by \citet{Trcka2022}. To this end, we apply the postprocessing method of \citet{Trcka2022} to the TNG100 simulation, the fiducial run of the IllustrisTNG simulation suite. Following \citet{Trcka2022}, we apply the MCRT code SKIRT (\citealt{Baes2011}; \citealp{Camps2015, Camps2020}) to a stellar mass-limited sample of $\approx$60\,000 TNG100 galaxies at $z=0$ and $z=0.1$ . We generate broadband fluxes in 53 broadband filters ranging from the GALEX far-UV (FUV) to the ALMA band 6 and low-resolution SEDs ranging from $0.1-2000\,\mu\mathrm{m}$ with $R=39$ for this sample of TNG100 galaxies.

To reveal potential biases in our postprocessing method and to further assess the realism of the cosmological simulation in the observational realm, we also explore different galaxy flux-flux and color-color relations over a large wavelength range. Since these relations trace the underlying distributions and scaling relations of physical properties (e.g. specific star-formation rate, dust mass, age), they provide an important testbed for the cosmological simulation plus radiative transfer postprocessing approach. This provides a complementary approach of assessing the simulation's realism, as the simulations are typically evaluated for their ability to reproduce the physical properties of the galaxy population inferred from observations (e.g. \citealt{Dave2017}; \citealt{DeRossi2017}; \citealt{Torrey2019}; \citealt{Rosito2019}; \citealt{Nelson2021}).

The outline of this paper is as follows: We describe the cosmological simulation as well as the SKIRT postprocessing method in Section~\ref{sec:Methods}, and compare TNG100 LFs to observations in Section~\ref{sec:Luminosity functions}. We proceed by comparing the simulated fluxes to observational data from the GAMA survey in terms of flux-flux and color-color relations (Section~\ref{sec:GAMA comparison}), and summarize our results in Section~\ref{sec:Conclusions}. We adopt a flat $\Lambda$CDM cosmology, with parameters measured by the Planck satellite (\citealt{Planck2016}), consistent with the IllustrisTNG cosmology. We use the AB magnitude system (\citealt{Oke1971}) throughout this study.

\section{Simulation methods}\label{sec:Methods}

\subsection{IllustrisTNG}

The IllustrisTNG suite (\citealt{TNG_Pillepich}; \citealt{TNG_Springel}; \citealt{TNG_Nelson}; \citealt{TNG_Naiman}; \citealt{TNG_Marinacci}) is a set of cosmological, magnetohydrodynamical simulations run using the moving-mesh code AREPO (\citealt{Springel2010}). The simulation suite consists of three different volumes with box sizes of approximately 50, 100, and 300 comoving Mpc, each realized with three to four different resolutions. All of these simulations were run with the same physical model, with the subgrid parameters chosen for the fiducial TNG100-1 run (hereafter `TNG100'), which is the highest-resolution run for the 100-cMpc box. Unlike in the EAGLE suite (\citealt{Schaye2015}), the subgrid parameters were not recalibrated for other IllustrisTNG simulations (at different resolutions and box sizes). For the cosmological parameters, the simulations use the 2015 results measured by the Planck satellite (\citealt{Planck2016}), i.e. $\Omega_m=0.3089$, $\Omega_b=0.0486$, $\Omega_\Lambda=0.6911$, $H_0=100\,h\,\mathrm{km\,s^{-1}Mpc^{-1}}$ with $h=0.6774$). In the following, we briefly describe the aspects of IllustrisTNG and its galaxy formation model (\citealt{Weinberger2017}; \citealt{Pillepich2018}) that are most relevant to this study.

TNG100 simulates a cube with box size of 110.7 comoving Mpc from $z=127$ to $z=0$. This volume is resolved with $1820^3$ baryonic and dark matter particles, corresponding to a mean particle mass of $1.4\times10^6\,M_\odot$ and $7.5\times10^6\,M_\odot$, respectively. Galaxies are identified as gravitationally bound substructures using the SUBFIND algorithm (\citealt{Springel2001}). Since molecular clouds cannot be resolved in the simulation, star formation is modelled stochastically for gas with $n_\mathrm{H}>0.106\,\mathrm{cm}^{-3}$ according to the two-phase model of \citet{Springel2003}. Stellar populations are modelled with a Chabrier initial mass function (\citealt{Chabrier2003}). These star particles subsequently affect the surrounding interstellar medium (ISM) via metal enrichment as well as feedback from supernovae explosions. The IllustrisTNG model furthermore incorporates gas radiative processes (including metal-line cooling and heating in an evolving UV background), formation and merging of supermassive black holes, as well as feedback from active galactic nuclei in a thermal and a kinetic mode.

In \citet{Trcka2022}, we calculated broadband fluxes (in 53 filters) as well as low-resolution SEDs ($R=39$) from $0.1-2000\,\mu\mathrm{m}$ for the TNG50-1 and the lower-resolution TNG50-2 simulations (\citealt{Nelson2019b}; \citealt{Pillepich2019}, see Table~\ref{tab:TNGruns} for an overview of the different simulation resolutions) and publicly released them on the IllustrisTNG website\footnote{\url{https://www.tng-project.org/}}. This data is available at two snapshots (099 and 091), corresponding to $z=0$ and $z=0.1$, for all galaxies above a stellar mass threshold. The stellar mass threshold of $10^8\,\mathrm{M}_\odot$ ensures that the galaxies are resolved by enough ($\gtrsim10^2$) star particles for the radiative transfer postprocessing. We remark that we always use the stellar mass within two stellar half-mass radii for the simulation stellar masses (as opposed to the total graviationally bound stellar mass for instance), which is available from the IllustrisTNG galaxy catalogue. 

With the present study, we add the same data products (i.e. broadband fluxes and low-resolution SEDs) for TNG100 at redshifts $z=0$ and $z=0.1$ for 61\,076 galaxies with $M_\star>10^{8.5}\,\mathrm{M}_\odot$ (we choose a higher stellar mass threshold for TNG100 due to the lower particle mass resolution compared to the TNG50 runs) to the database. For the galaxy samples of all three simulations, subhalos that are flagged as being not of cosmological origin are excluded\footnote{The IllustrisTNG subhalo finder sometimes falsely identifies baryonic fragments or clumps as galaxies. The IllustrisTNG galaxy catalogue (\citealt{Nelson2019a}) contains a flag that indicates if a subhalo is probably not of cosmological origin, in which case the `SubhaloFlag' field is set to zero. We omit these objects from the postprocessing analysis.}. An overview of the different sample definitions and galaxy sample sizes is shown in Table~\ref{tab:TNGruns}. We caution that the chosen stellar mass thresholds are relatively low, meaning that the postprocessing results for the lowest-mass galaxies could be unreliable for TNG100-1 and TNG50-2.

\begin{table}
    \centering
    \begin{tabular}{cccccc}
         Simulation & $V\,[\mathrm{cMpc}^3]$ & $m_\mathrm{b}\,[\mathrm{M}_\odot]$ & $M_\star^\mathrm{min}\,[\mathrm{M}_\odot]$ & $N_\mathrm{gal}^{z=0}$ & $N_\mathrm{gal}^{z=0.1}$\\ \hline
        TNG100-1 & $106.5^3$ & $1.4\times10^6$ & $10^{8.5}$ & 30\,712 & 30\,364\\
        TNG50-1 & $51.7^3$ & $8.5\times10^4$ & $10^8$ & 7\,375 & 7\,302\\
        TNG50-2 & $51.7^3$ & $6.8\times10^5$ & $10^8$ & 5\,669 & 5\,665
  
    \end{tabular}
    \caption{Runs of the IllustrisTNG suite that we consider in this study. For each simulation, we list the volume, the target baryon mass (the resolution) and the stellar mass (more specifically, the stellar mass in two stellar half-mass radii) threshold which defines the galaxy samples. $N_\mathrm{gal}$ indicates the number of galaxies (in the snapshots at $z=0$ and $z=0.1$) that conform to our sample selection criteria.}
    \label{tab:TNGruns}
\end{table}

\subsection{Radiative transfer postprocessing}

The methodology for the radiative transfer postprocessing adopted here for TNG100 galaxies is exactly the same as in \citet{Trcka2022}, which, in turn, is based on \cite{Camps2016, Camps2018} and \citet{Kapoor2021}. We briefly summarize the main steps here and refer the reader to \citet{Trcka2022} for more details.

We use the 3D dust MCRT code SKIRT (\citealt{Baes2011}; \citealp{Camps2015, Camps2020}) to generate broadband fluxes over a large (UV-FIR) wavelength range. We simulate the emission of photon packets from evolved stellar populations as well as star-forming regions (`primary emission' in SKIRT). The photon packets are then propagated through the dusty ISM, where they get absorbed and scattered. Furthermore, the dust grains are stochastically heated and subsequently emit IR radiation (`secondary emission', \citealt{Camps2015b}). Finally, the photon packets are recorded in synthetic instruments that emulate different apertures, orientations, and broadband filters. We briefly describe the different components of the SKIRT simulations and how they are imported from IllustrisTNG in the following.

\begin{itemize}

    \item Evolved stellar populations: All star particles with ages above 10\,Myr are treated as evolved stellar populations. We model their SED using the \citet{Bruzual2003} template library with a Chabrier IMF. All parameters to model the emission of evolved stellar populations (positions, current masses, metallicites, ages, and smoothing lengths) are directly available from the IllustrisTNG snapshot data.

    \item Star-forming regions: Star particles with ages below 10\,Myr are modelled as star-forming regions, i.e. young stars that are still partially enshrouded within their dusty birth clouds. We use the template library MAPPINGS-III (\citealt{Groves2008}) to model their SED, which contains the light contribution from the young stellar population as well as nebular and dust emission. In addition to the positions, metallicities, and smoothing lengths, this template library has a number of parameters that are not directly available from the snapshot data. These are the star-formation rates (calculated as initial mass of the star particle divided by its age), ISM pressure (set to a constant value of $P/k_B=10^5\,\mathrm{K\,cm^{-3}}$), and compactness parameter (randomly sampled from a Gaussian distribution). Lastly, the photodissociation region (PDR) covering factor is calculated as $f_\mathrm{PDR}=e^{-t/\tau}$, with $t$ being the age of the star particle and $\tau$ a free parameter in the radiative transfer postprocessing scheme. 

    \item Diffuse dust: As IllustrisTNG does not track the dust content in the ISM, we assign dust to gas cells based on their metallicity. Specifically, we use the criterion of \cite{Torrey2012, Torrey2019} to select dust-containing gas cells based on their temperature and mass density. This criterion separates the hot circumgalactic medium (CGM) from the ISM. While we do not assign dust to the CGM gas cells, the dust mass in all other cells is scaled to their metal masses, with the dust-to-metal ratio $f_\mathrm{dust}$ being a free parameter of the postprocessing scheme. All other parameters that control the diffuse dust (positions, mass densities, temperatures, and metallicities) are directly available from the snapshot data. For the optical properties of the diffuse dust, we use the THEMIS dust model from \citet{Jones2017}. The dusty medium is discretised on an octtree grid (\citealt{Saftly2013,Saftly2014}) with a maximum subdivision level of twelve.
    
\end{itemize}

The SKIRT postprocessing simulations are performed for a defined spatial domain. In our case, we use a cube with side length ten times stellar half-mass radii, centered on the subhalo positions. Additionally, we consider only star particles within a sphere of radius five stellar half-mass radii\footnote{This ensures that we capture most of the starlight emitted by the galaxy. To test this more quantitatively, we compared half-light sizes derived by \citet{Baes2024b} for massive ($M_\star\geq10^{9.8}\,\mathrm{M}_\odot$) TNG50-1 galaxies to their half-mass sizes. The bluest available band (LSST u) shows the highest half-light to half-mass size ratios, with $28.3\,\%$ ($3.47\,\%$) of all galaxies having a half-light size larger than two (five) half-mass radii. Hence, there is a sizeable fraction of galaxies for which we miss some starlight in the bluest optical and the UV filters, but we remark that our maximum aperture of five stellar half-mass radii is comparable or larger than the observational apertures used in this paper (see Figure~\ref{fig:apertures}).} for the postprocessing. Lastly, we use $5\times10^7$ photon packets to perform the radiative transfer simulations.

In \citet{Trcka2022}, the free parameters $\tau$ and $f_\mathrm{dust}$ were calibrated using a test sample of TNG50 galaxies which are compared to low-redshift multiwavelength observational data from the DustPedia archive\footnote{\url{http://dustpedia.astro.noa.gr/}} (\citealt{Davies2017}; \citealt{Clark2018}). Using various luminosity and color scaling relations, the default parameters were determined to $\tau=3\,\mathrm{Myr}$ and $f_\mathrm{dust}=0.2$. We kept these parameters unchanged for the postprocessing of TNG100 galaxies. We have verified that the TNG100 galaxies exhibit a similar behaviour compared to TNG50 on the scaling relations that were used to calibrate the free parameters.

\begin{figure*}
    \centering
    \includegraphics[width=\textwidth]{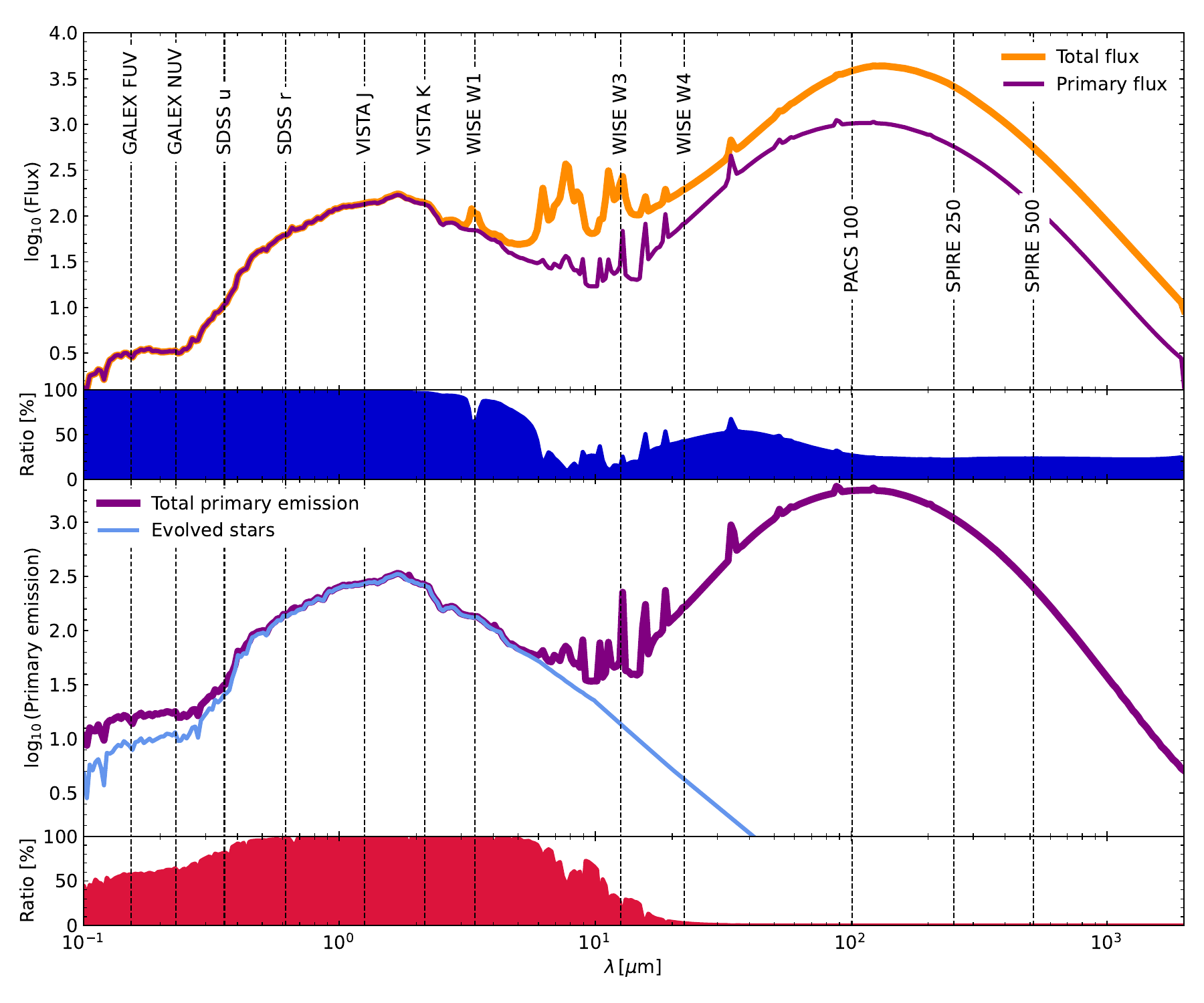}
    \caption{The global SED of the TNG100 galaxy population, split into the various components of our SKIRT simulations. The dashed black lines correspond to the pivot wavelengths of the broadband filters that we consider in Section~\ref{sec:GAMA comparison}. The first panel shows the total SED (orange line) as well as the SED from the attenuated primary emission (evolved stellar populations and star-forming regions, purple line). The ratio between the two is indicated by the blue area in the second panel. The third panel shows the unattenuated primary emission (purple line) and the contribution from evolved stellar populations (light blue line), with the ratio of the two indicated by the red area (fourth panel). Note that the SEDs in the first and third panels are shown on arbitrarily normalized logarithmic scales.}
    \label{fig:SEDbreakdown}
\end{figure*}

To illustrate our methodology, we show the global SEDs of the TNG100 galaxy population in Figure~\ref{fig:SEDbreakdown} split into the different components of our SKIRT simulations. These global SEDs are obtained by summing the rest-frame SEDs of all TNG100 galaxies in our base sample at $z=0$ and $z=0.1$. The total SED is shown by the orange line in the first panel of Figure~\ref{fig:SEDbreakdown}. To separate primary emission (from evolved stars and star-forming regions) and secondary emission (from diffuse dust in the ISM), we show the SED arising solely from primary emission (which is still attenuated by the diffuse dust) by the purple line in the first panel of Figure~\ref{fig:SEDbreakdown}. The ratio of primary to total flux is indicated by the filled blue area, which is shown on a linear scale in the second panel.

To further split the primary emission into the contribution from evolved stellar populations and star-forming regions, we show the unattenuated primary emission of the galaxy population by the purple line in the third panel of Figure~\ref{fig:SEDbreakdown}. The emission considering only evolved stellar populations is shown in light blue, while the filled red area indicates the fractional contribution of evolved stellar populations to the total primary emission. Figure~\ref{fig:SEDbreakdown} exhibits the expected features of the different components of our SKIRT setup: In the optical and NIR the flux is dominated by evolved stellar populations, while at larger wavelengths diffuse dust emission dominates. Star-forming regions contribute significantly to the UV and also partially to the MIR and FIR flux due to the emission from their dusty birth clouds.

\subsection{Simulation products}

The main output of the radiative transfer postprocessing are broadband fluxes in 53 filters, from the UV (GALEX FUV) to the ALMA band 6. These fluxes are available for all galaxies in TNG100-1 (as well as TNG50-1 and TNG50-2 already presented by \citealt{Trcka2022}) above the stellar mass threshold (see Table~\ref{tab:TNGruns}), at redshifts 0 and 0.1. The broadband flux is given both in the galaxy rest-frame (in absolute AB magnitudes) and in the observational frame\footnote{For the data in the observational frame, the SKIRT instrument is placed at 20\,Mpc for redshift zero or at the corresponding redshift for $z=0.1$.} (in Jy). Additionally, we provide low-resolution SEDs ($R=39$) in the observational frame (in Jy) for $0.1-2000\,\mu\mathrm{m}$ for all TNG100-1 galaxies in the base sample. All data are available in three different galaxy orientations (random, edge-on, and face-on) as well as four different circular apertures (with aperture radii of 10\,kpc, 30\,kpc, two stellar half-mass radii, and five stellar half-mass radii).

\section{Galaxy luminosity functions}\label{sec:Luminosity functions}

We begin by investigating low-redshift luminosity functions in various broadband filters. As in \citet{Trcka2022}, we use the rest-frame magnitudes (which we convert into solar luminosities) for our main galaxy sample which combines the $z=0$ and $z=0.1$ snapshots. We use a default orientation (random) throughout this work and adopt an aperture of five stellar half-mass radii in Section~\ref{sec:Luminosity functions}, the default choice for the simulated LFs in \citet{Trcka2022}. Since the observational LFs are thought to be representative of the local galaxy population, we do not mimic any observational selection effect (as instead done and described in Section~\ref{sec:Sensitivity}). 

In \citet{Trcka2022} (their figure 9), luminosity functions of the TNG50-1 simulation were found to overestimate the observational estimates in all filters and at all luminosities from the UV to the far-IR. At the bright end, this discrepancy can be mitigated by choosing a significantly smaller aperture (10\,kpc instead of the default five stellar half-mass radii), but this value is less representative of observational apertures and does not resolve the tension for galaxies fainter than the knee of the luminosity functions. \citet{Trcka2022} found that the discrepancy is largely mitigated when using the lower-resolution TNG50-2 simulation. Indeed, within the IllustrisTNG model, the resolution improvement from the fiducial TNG100 resolution to TNG50 results in somewhat larger galaxy masses and SFRs (\citealt{Pillepich2018,TNG_Pillepich,Pillepich2019}; \citealt{Donnari2019}). We test this statement here explicitly by investigating the LFs of the TNG100 simulation, which is the fiducial resolution at which the subgrid parameters were chosen.

\begin{figure*}
    \centering
    \includegraphics[width=\textwidth]{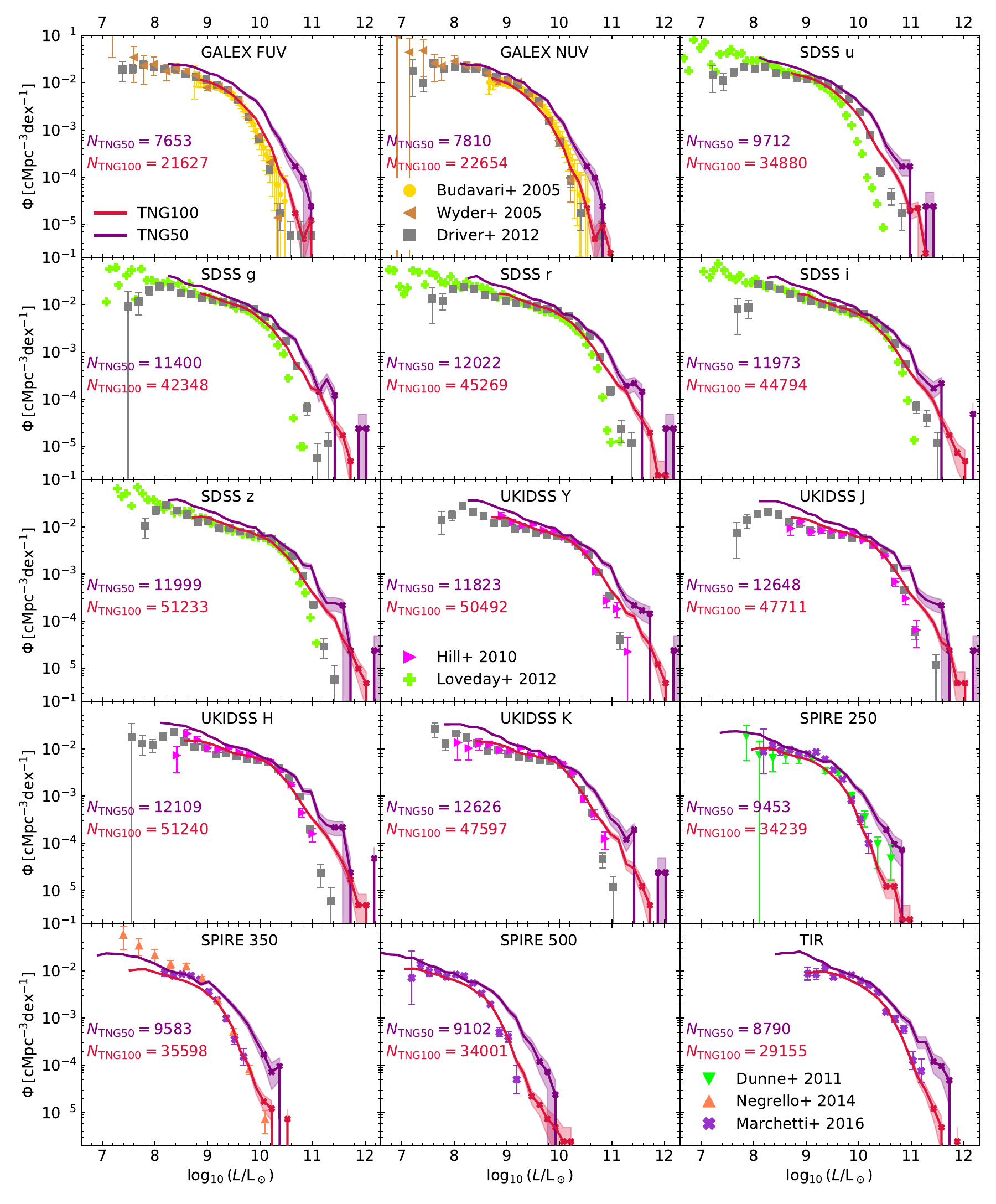}
    \caption{Luminosity functions in 14 bands and the total infrared (TIR). Continuous lines mark the simulation results for TNG50 (blue) and TNG100 (red), for $z\le0.1$. The shaded area corresponds to the Poisson error, crosses mark luminosity bins with fewer than ten galaxies. The simulated luminosity functions are computed only above a completeness limit, see text for details. The number of simulated galaxies above this completeness limit is shown in each panel. Observational data are shown as various markers. The TNG100 LFs are in excellent agreement with the observations.}
    \label{fig:LFs}
\end{figure*}

We show the low-redshift luminosity functions for TNG100 in Figure~\ref{fig:LFs}. The observational estimates from various low-redshift surveys\footnote{Specifically, we use LF data at $z\lesssim0.1$ from the GALEX MIS (\citealt{Budavari2005}), GALEX AIS (\citealt{Wyder2005}), GAMA (\citealt{Driver2012}), SDSS (\citealt{Loveday2012}), SDSS + UKIDSS LAS + MGC redshift survey (\citealt{Hill2010}), H-ATLAS (\citealt{Dunne2011}), \textit{Planck} ERCSC (\citealt{Negrello2014}), and \textit{Spitzer} Data Fusion database + HerMES (\citealt{Marchetti2016}) surveys.} are equivalent to the ones from \citet{Trcka2022} (see their section 3.2.1 for more details), which are corrected to $h=0.6774$ to be consistent with the cosmological parameters of IllustrisTNG. We also include the LFs from the TNG50-1 (hereafter `TNG50') simulation to highlight the convergence behaviour of the cosmological simulations. To not overcrowd the figure TNG50-2 is not shown, but we note that the TNG50-2 LFs closely align with the TNG100-1 results, meaning that the LFs are converged with simulation box size. In Figure~\ref{fig:LFs}, the Poisson error for the simulated LFs is shown as shaded area, and luminosity bins with fewer than ten galaxies are marked. We cut the calculation of the simulated LFs at a minimum luminosity to ensure that the shown LFs are complete. This minimum luminosity is calculated as the 90\,\% luminosity percentile in the lowest 5\,\% stellar mass bin. The number of galaxies above this luminosity threshold are noted in each panel for TNG100 and TNG50 separately.

Figure~\ref{fig:LFs} shows how the agreement between TNG100 and observational LFs improves compared to TNG50. In fact, the TNG100 LFs provide an excellent match to the observational data in the near-UV (NUV) and FIR bands. In the FUV, optical and near-infrared (NIR) bands (GALEX FUV, SDSS and UKIDSS filters), the faint ends and knees of the observed LFs are also precisely reproduced in TNG100. At the bright ends TNG100 overestimates the observational estimates, but we note that in this regime there are also large differences across the observational datasets. As an example, the LFs in the SDSS filters from \citet{Loveday2012} are given in Petrosian apertures, while \citet{Driver2012} use Kron apertures. Even though both studies use data from the GAMA survey, the differences in the LFs reach almost an order of magnitude for the brightest luminosity bins (see also \citealt{Hill2011} and \citealt{Bernardi2013} for a discussion on this issue). For a detailed discussion on the impact of the aperture for the simulated LFs, we refer the reader to section 4 in \citet{Trcka2022}.

We conclude that, as suggested by \citet{Trcka2022}, the way that the subgrid parameters are chosen in the IllustrisTNG model (at the fiducial TNG100 resolution) indeed caused the discrepancy in the LFs for TNG50. Acknowledging observational uncertainties at the bright end related to aperture choices, the agreement between TNG100 and low-redshift observational LFs is excellent.

\section{UV-submm broadband fluxes: Comparison with GAMA}\label{sec:GAMA comparison}

To assess galaxy scaling relations and distributions in the observational realm, we continue by analyzing different flux-flux and color-color relations over a large wavelength range. As opposed to analyzing scaling relations in the physical realm, this analysis provides a complementary approach of assessing the simulation's realism. We also use these relations to evaluate the accuracy and reveal potential systematics in our radiative transfer postprocessing scheme. We only analyze TNG100 in this section, and refer the reader to Appendix~\ref{app:resolution} for a comparison to TNG50.

We first detail the observational dataset in Section~\ref{sec:GAMA data} and describe how we homogenize the observational and simulated galaxy samples in Section~\ref{sec:Sensitivity}, before discussing the results for the flux-flux and color-color relations in Sections~\ref{sec:Flux-flux relations} and~\ref{sec:Color-color relations}, respectively.

\subsection{Observational data from GAMA}\label{sec:GAMA data}

The Galaxy and Mass Assembly (GAMA) survey (\citealp{Driver2009, Driver2011}; \citealt{Liske2015}; \citealt{Baldry2018}; \citealt{Driver2022}) is a spectroscopic survey of galaxies with the AAOmega spectrograph in the optical wavelength range, mounted on the Anglo Australian Telescope (AAT). The survey consists of five different fields with varying input catalogues (used for target selection), observing a total area of 286\,deg$^2$. The most recent data release (DR4) of GAMA (\citealt{Driver2022}) contains spectra, spectroscopic redshifts, X-ray to FIR photometry from various other surveys\footnote{Specifically, the GAMA database includes photometry from the XMM-XXL, GALEX, SDSS, KiDS, VIKING, WISE, and Herschel-ATLAS surveys.}, as well as derived data such as stellar masses and rest-frame fluxes for some $\sim300\,000$ galaxies. All accessed data used in this study is part of the GAMA data release 4, described in \citet{Driver2022}. Due to its large sample size of low-redshift galaxies ($z\lesssim0.6$) and large wavelength coverage of photometric data, the GAMA database provides an excellent observational sample to compare to the simulated photometric data from TNG100.

The GAMA project consists of three phases, which are different in their target selection as the input catalogues were updated with more recent photometric data from other surveys over time. As the three equatorial fields (labelled G09, G12, and G15) observed as part of GAMA II have the highest availability of derived data products (importantly, those are the only galaxies within GAMA with matched-aperture photometry), we only use this dataset throughout this study. The target selection is defined as having an apparent Petrosian $r$-band magnitude below $19.8\,\mathrm{mag}$ in SDSS DR7. This limit is the same for all three fields.

For the analysis in this paper, we use various catalogues from the GAMA database, which we describe in this section. To select only galaxies that are part of the main GAMA II survey, we use the TilingCat v46 catalogue from the EqInputCat data management unit (DMU). Objects that are part of the main survey have a survey class of four or higher. We enforce this criterion for our GAMA sample.

We use broadband fluxes from the LambdarCat v01 catalogue in the LambdarPhotometry DMU. In this catalogue, the fluxes are extracted using matched aperture photometry with the \textit{Lambdar} code (\citealt{Wright2016}). \textit{Lambdar} measures the photometry given an input aperture (in this case, the apertures come from a combination of SExtractor (\citealt{Bertin1996}) runs on the SDSS $r$ and VIKING $Z$-bands of imaging as well as visual inspection) and performs aperture convolution, deblending, correction, and sky substraction. The fluxes are available for the GALEX, SDSS, VISTA, WISE, PACS, and SPIRE bands. The fluxes are corrected for Milky Way extinction but not K-corrected, hence these are fluxes in the observational frame (as opposed to rest-frame fluxes).

As we want to limit the observational galaxy sample in redshift, we also download redshift estimates from the DistanceFrames v14 catalogue in the LocalFlowCorrection DMU (\citealt{Baldry2012}). We use the redshifts from the Tonry flow model (\citealt{Tonry2000}), which equals the cosmic microwave background redshift for $z\geq0.03$ and takes into account local flows at lower redshifts. Following the documentation of this DMU, we impose $z\geq0.002$ as lower-redshift objects are potentially not galaxies. We also impose $z\leq0.1$ to not extrapolate our simulation results into higher redshift ranges. Only galaxies with a high-quality redshift (redshift flag must be three or larger) are kept in our sample. 

We also impose a stellar mass limit ($M_\star\geq10^{8.5}\,\mathrm{M}_\odot$) to the GAMA galaxies, the same stellar mass limit of our TNG100 galaxy sample. Stellar masses\footnote{Different stellar mass estimates exist in this GAMA table. While the TNG100 stellar masses would correspond to the sum of the GAMA stellar and remnant masses, we just consider the more commonly used stellar masses. Adding the remnant masses would shift the stellar masses by less than 0.1 dex. Furthermore, we correct the stellar masses to $h=0.6774$, but do not perform an aperture correction.} are obtained from the StellarMassesLambdar v20 catalogue in the StellarMasses DMU. These are inferred from SED fits to the \textit{Lambdar} aperture photometry (see \citealt{Taylor2011} for details). The cuts in survey class ($\geq4$), redshift flag ($\geq3$), redshift ($0.002\leq z\leq0.1$), and stellar mass ($M_\star\geq10^{8.5}\,\mathrm{M}_\odot$) lead to a base sample of 17\,932 galaxies contained in the GAMA dataset. We note that not all galaxies in this GAMA catalogue have detected broadband fluxes in all filters.

The GAMA base sample is then cut further depending on the broadbands that are involved in a specific flux-flux or color-color plot. We first impose SNR cuts on all involved filters to ensure that the GAMA galaxies have reliable fluxes. Specifically, we discard all galaxies with $\mathrm{SNR}<3$ in any of the involved filters. In a second step, we want to define a flux threshold that broadly corresponds to a volume-limited sample. The same threshold can then be applied to the simulated galaxies to ensure a fair comparison. We noted that the GAMA galaxies exhibit noise distributions with outliers multiple orders of magnitude below the median, even after this SNR cut. This leads to some galaxies having very low flux values, which are not representative of the typical sensitivity of the respective surveys. Hence, we compute the 10\,\%-percentiles of the GAMA galaxies with $\mathrm{SNR}>3$ in each band, and use these fluxes as thresholds for the GAMA and TNG100 datasets. This means that in every flux-flux or color-color plot, if a GAMA or TNG100 galaxy has a flux below the threshold in any band it is omitted from the plot. The flux thresholds are given in Table~\ref{tab:FluxLimits} for all filters considered in Figures~\ref{fig:fluxes} and~\ref{fig:colors}.

We caution that the choice of SNR (3) and flux (10\,\%-percentile) thresholds are arbitrary. We have tested different strategies (changing the SNR and flux percentile values, or either just using an SNR or a flux percentile criterion), and find that the peaks and correlations of the distributions are hardly affected. On the other hand, the widths of the distributions are altered (e.g. lowering or dropping the SNR criterion primarily makes the GAMA distribution wider). We adopted the specific thresholds as a compromise between galaxy sample size and mitigating noise and incompleteness effects in GAMA. For our chosen thresholds, we find the GAMA noise levels moderate in the sense that the widths of the flux and color distributions of TNG100 and GAMA are similar, i.e. the intrinsic scatter in the galaxy population dominates over instrumental effects. Due to this ambiguity in SNR and flux thresholds, we focus the discussion in Sections~\ref{sec:Flux-flux relations} and~\ref{sec:Color-color relations} on the peaks and correlations of the distributions. Making firm statements about the scatter of the shown flux-flux and color-color relations would require adding realistic GAMA-like noise to the TNG100 galaxies, which is beyond the scope of this study.

\begin{figure}
    \centering
    \includegraphics[width=\columnwidth]{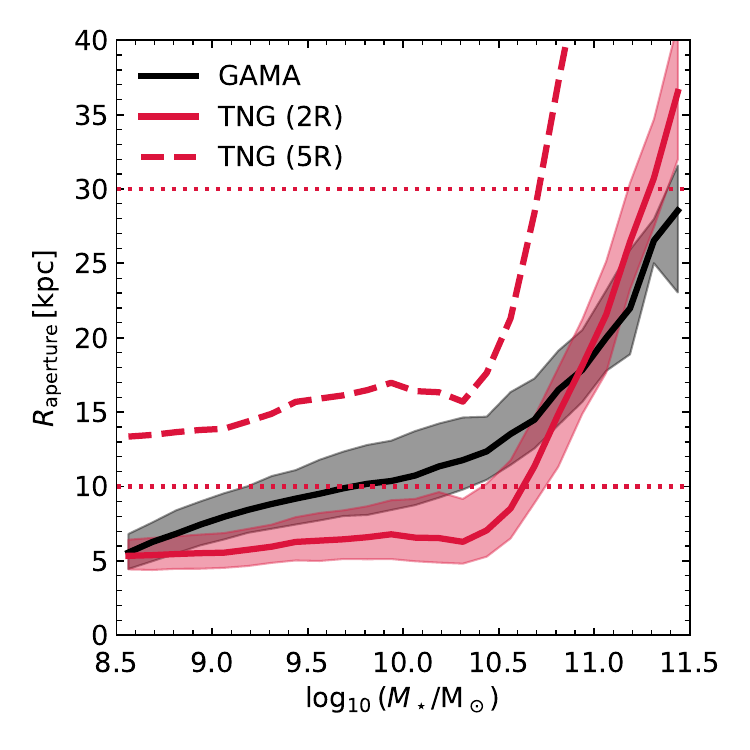}
    \caption{Apertures of TNG100 galaxies (red) and objects from the GAMA survey, for $0.002\leq z\leq0.1$ and $M_\star\geq10^{8.5}\,\mathrm{M}_\odot$. The GAMA apertures correspond to the cricularized radii of the elliptical \textit{Lambdar} apertures. For TNG100 we show the constant apertures (10 or 30\,kpc) as dotted lines. The other available TNG100 apertures (2 or 5 stellar half-mass radii) and the GAMA apertures are displayed as running medians as a function of stellar mass. Shaded areas indicate the interquartile range (not shown for 5 stellar half-mass radii). Since an aperture of two stellar half-mass radii provides the closest match to the GAMA apertures, we adopt this as our default aperture for the TNG100 fluxes in Section~\ref{sec:GAMA comparison}.}
    \label{fig:apertures}
\end{figure}

Lastly, we remark that aperture mismatches in the observed and simulated datasets can substantially bias the comparison. The distribution of GAMA apertures (given as the circularized radii\footnote{We use $R_\mathrm{aperture}=\sqrt{ab}$ with $a$ and $b$ the semi-major and semi-minor axes of the aperture, respectively.} of the elliptical aperture used by \textit{Lambdar}) as a function of stellar mass for all 17\,932 galaxies in the base sample is shown in Figure~\ref{fig:apertures}. These apertures are compared to the four different available apertures for the TNG100 data (10\,kpc, 30\,kpc, 2 or 5 stellar half-mass radii). We find that two stellar half-mass radii provide the closest match to the GAMA apertures, even though the TNG100 apertures are significantly smaller for all stellar masses below $10^{11}\,\mathrm{M}_\odot$ in that case. Hence, we adopt two stellar half-mass radii as our default aperture in Section~\ref{sec:GAMA comparison}.

\subsection{Observational sensitivity limits for simulated galaxies}\label{sec:Sensitivity}

A major caveat when comparing observational and simulated datasets is that the galaxy samples can be very different. This caveat is usually mitigated by matching the samples in some physical properties like stellar masses or star-formation rates (e.g. \citealt{Diemer2019}; \citealt{Donnari2021b}; \citealt{Trcka2022}; \citealt{Goddy2023}). However, this approach bears the problem that the observational and simulated definitions of those properties can be different\footnote{As an example, the galaxy star-formation rate in the simulation is typically defined as the instantaneous SFR of the star-forming gas. On the other hand, in observations the SFR is determined for some tracer of young stellar populations, yielding the average SFR over a certain timescale.}, and physical parameters inferred from observations come with their own caveats. Hence, we implement a different method to homogenize the galaxy samples.

We base our method on the observational sensitivity limits in various filters, which determine the flux limits of the galaxies. We use these limits to filter out `fainter' TNG100 galaxies which would lie below the observational detection threshold. This approach is similar to postprocessing studies of semi-analytical models (SAMs) over large redshift ranges which have been used to study galaxy clustering (\citealt{Blaizot2005}; \citealt{Kitzbichler2007}). In this approach, the (periodic) simulation box at different snapshots is stacked many times to construct a sufficiently large volume and to calculate a mock lightcone. Unfortunately, such a mock lightcone construction requires the postprocessing of many different snapshots, which is feasible for the SAM postprocessing but prohibitive for our 3D dust radiative transfer modelling. Hence, we do not stack the simulation box at different snapshots, but rather place the friend-of-friend halos (FoF groups) of the $z=0$ and $z=0.1$ snapshots at arbitrary distances (within the redshift bounds from the observational sample, i.e. $0.002<z<0.1$) from the mock observer. We assume that the halos are uniformly distributed in space, such that the comoving number density of halos $n$ is constant:

\begin{equation}
    n(D_c)=\frac{N(D_c, D_c+\mathrm{d}D_c)}{V(D_c, D_c+\mathrm{d}D_c)}=\text{const}=\frac{N_\mathrm{tot}}{V_\mathrm{tot}}.
\end{equation}
Here, $N_\mathrm{tot}$ denotes the total number of halos from TNG100 that are now distributed, $N(D_c,D_c+\mathrm{d}D_c)$ indicates the number of halos within a small comoving distance interval $\mathrm{d}D_c$, and $V(D_c,D_c+\mathrm{d}D_c)$ corresponds to the volume of this comoving distance slice. The total comoving volume of the (mock) survey, $V_\mathrm{tot}$, is given by the redshift limits $z_\mathrm{min}$ and $z_\mathrm{max}$:

\begin{equation}
    V_\mathrm{tot}=\frac{4\pi}{3}\bigl(D_c(z_\mathrm{max})^3-D_c(z_\mathrm{min})^3\bigr).
\end{equation}
The normalized probability distribution function for `placing' a halo at a specific distance, $p(D_c)$, can then be written as follows:

\begin{equation}
    p(D_c)\mathrm{d}D_c=\frac{N(D_c,D_c+\mathrm{d}D_c)}{N_\mathrm{tot}}=\frac{4\pi D_c^2\mathrm{d}D_c}{V_\mathrm{tot}}.\end{equation}

With this procedure, we draw random redshifts within $0.002\leq z\leq0.1$ for each TNG100 halo and then assign these random halo redshifts to all subhalos (i.e. galaxies) that belong to a particular halo. This is done independently for the $z=0$ and $z=0.1$ snapshots. We then compute the broadband flux $F_\nu^\mathrm{j}(z)$ in any filter j of the galaxy at that arbitrary redshift. Since we need this flux in the observational frame, we cannot simply use the fluxes that we stored for the TNG100 galaxies (they are stored in the rest- and in the observational frame, but only at the fixed snapshot redshifts of 0 and 0.1). Hence we convolve the low-resolution SED $F_\nu(z_\mathrm{snap}, \lambda)$ (which is stored for each galaxy in the observational frame at its snapshot redshift $z_\mathrm{snap}$) with filter transmission curves\footnote{We obtained the filter transmission curves from the Spanish Virtual Observatory (SVO) filter profile service (\url{http://svo2.cab.inta-csic.es/theory/fps/}). For photon} counter instruments, the transmission curves are multiplied by the wavelengths. $T^\mathrm{j}(\lambda)$, accounting for the redshifting of the photons:

\begin{equation}\label{eq:RedshiftedFlux}
    F_\nu^\mathrm{j}(z)=\frac{\int T^\mathrm{j}(\lambda\cdot k)\cdot F_\nu(z_\mathrm{snap},\lambda)\cdot k\,\mathrm{d}\lambda}{\int T^\mathrm{j}(\lambda\cdot k)\,\mathrm{d}\lambda}\times\frac{D_l(z_\mathrm{snap})^2}{D_l(z)^2},
\end{equation}
with $k=(1+z)/(1+z_\mathrm{snap})$ and $D_l$ indicates the luminosity distance (for the $z=0$ snapshot we use $D_l=20\,\mathrm{Mpc}$ as the SKIRT instrument is placed at this distance). Placing TNG100 galaxies at arbitrary redshifts introduces inconsistencies due to galaxy evolution between the snapshot redshift from which they were extracted and the new redshift at which they are placed. The unknown result without this systematic effect, which would be obtained if we had access to each galaxy at the random continuous redshift between 0.002 and 0.1, is bound by the results using only one of the $z=0$ and $z=0.1$ snapshots. To estimate if this inconsistency affects our results, we repeat our analysis using only the snapshot $z=0$ and $z=0.1$, respectively. We find that none of our results are affected significantly\footnote{The similarity of the TNG100 and GAMA distributions quantified by the 2D Kolmogorov-Smirnov test statistic $D_\mathrm{KS}$ never changes by more than 0.04 in Figures~\ref{fig:fluxes} and~\ref{fig:colors}.}.

The end product of this procedure are observer-frame fluxes in Jansky (Eq.~\ref{eq:RedshiftedFlux}) for the entire $z=0$ and $z=0.1$ TNG100 galaxy sample, in all available 53 filters. These fluxes can be computed for continuous redshifts within arbitrary redshift intervals, and readily used to mimic observational sensitivity limits in various filters. We emulate the observational galaxy selection by distributing the TNG100 galaxies over the same redshift range ($0.002<z<0.1$) as the GAMA dataset. Consistent with the GAMA data, only TNG100 galaxies with fluxes above the thresholds from Table~\ref{tab:FluxLimits} are shown in Figures~\ref{fig:fluxes} and~\ref{fig:colors}. Under the assumption that the GAMA data is complete (i.e. volume-limited) above these flux limits, this procedure mitigates any sample selection effects to ensure a fair comparison of the TNG100 and GAMA galaxy samples.

\begin{table}
    \centering
    \begin{tabular}{ccc}
         Filter & Pivot wavelength\,[$\mu$m] & Flux limit\,[Jy] \\ \hline
        GALEX FUV & 0.154 & $4.48\times10^{-6}$ \\
        GALEX NUV & 0.230 & $6.36\times10^{-6}$ \\
        SDSS $u$ & 0.356 & $1.41\times10^{-5}$ \\
        SDSS $r$ & 0.618 & $5.76\times10^{-5}$ \\
        VISTA $J$ & 1.25 & $9.42\times10^{-5}$ \\
        VISTA $K$ & 2.21 & $1.00\times10^{-4}$ \\
        WISE W1 & 3.39 & $1.26\times10^{-4}$ \\
        WISE W3 & 12.6 & $4.02\times10^{-4}$ \\
        WISE W4 & 22.3 & $\mathbf{4.25\times10^{-3}}$ \\
        PACS 100 & 101 & $5.39\times10^{-2}$ \\
        SPIRE 250 & 253 & $2.10\times10^{-2}$ \\
        SPIRE 500 & 515 & $2.15\times10^{-2}$
    \end{tabular}
    \caption{Flux limits for the various broadband filters used to construct flux-flux and color-color relations (Figures~\ref{fig:fluxes} and~\ref{fig:colors}). These flux limits correspond to the 10\,\%-percentile of the GAMA fluxes with $\mathrm{SNR}>3$ in each filter. Only galaxies (for both the GAMA and TNG100 samples) which have fluxes above these thresholds in all involved filters for a specific flux-flux/color-color relation are plotted in this relation in Figures~\ref{fig:fluxes} and~\ref{fig:colors}.}
    \label{tab:FluxLimits}
\end{table}

\subsection{Galaxy flux-flux relations}\label{sec:Flux-flux relations}

We compare the simulated and observed fluxes in six different flux-flux relations in Figure~\ref{fig:fluxes}. We always consider the VISTA $K$ band in combination with various other bands. The $K$ band is a good tracer for stellar mass (\citealt{Kauffmann1998}; \citealt{Bell2001}), hence this analysis is analogous to various galaxy scaling relations as a function of stellar mass in the `observational realm'. In Figures~\ref{fig:fluxes} and~\ref{fig:colors}, we show the GAMA and TNG100 2D distributions as kernel density estimates (KDE), with the contours indicating various percentiles of enclosed fraction of the galaxy population density. 1D histograms are shown on the sides, and observational errors are indicated by the grey ellipses in the upper left corner where the darker (lighter) ellipse indicates the median (upper quartile) 1-$\sigma$ observational error bar. The error bars are computed by propagating the flux uncertainties in quadrature, assuming that the flux uncertainties are uncorrelated with each other. The number of galaxies above the flux limits are given both for TNG100 and GAMA in the top right of each panel. To quantify the degree of agreement between the TNG100 and GAMA distributions, we also compute the two-dimensional Kolmogorov-Smirnov test statistic $D_\mathrm{KS}$ (\citealt{Kolmogorov1933}; \citealt{Smirnov1948}). This number is given in the top right of each panel, with lower numbers indicating a better agreement between the two distributions.

We also discuss two alternative realizations of Figure~\ref{fig:fluxes} in the appendix. Figure~\ref{fig:FluxesTNG50} displays the exact same flux-flux relations, but using TNG50 instead of TNG100 to explore the impact of the simulation resolution. In Figure~\ref{fig:conditionalKDE}, we test the same flux-flux relations using a conditional KDE, i.e. exactly matching the TNG100 and GAMA VISTA $K$ distributions.

All results in Figures~\ref{fig:fluxes} and~\ref{fig:colors} use a TNG100 aperture of two stellar half-mass radii, which is systematically smaller than the GAMA apertures (see Figure~\ref{fig:apertures}). To verify if this aperture choice significantly affects our results, we have reproduced (but do not show) all flux-flux and color-color relations using a TNG100 aperture of five stellar half-mass radii. We find that the differences are minor ($D_\mathrm{KS}$ never changes by more than 0.05) and do not affect any of our conclusions.

\subsubsection{VISTA K vs. GALEX FUV}\label{sec:VISTA K vs. GALEX FUV}

The relation between galaxy stellar mass and star-formation rate is a fundamental galaxy evolution diagnostic (e.g. \citealt{Popesso2023}). We begin the TNG100-GAMA comparison by showing an analogue of this fundamental relation in the observational realm: VISTA $K$ versus GALEX FUV luminosity (top left panel of Figure~\ref{fig:fluxes}). The FUV-luminosity is dominated by young stellar populations and hence traces SFR (modulo dust attenuation effects, e.g. \citealt{Salim2007}).

The TNG100 and GAMA distributions in this flux-flux relation match to excellent precision ($D_\mathrm{KS}=0.08$), with both datasets showing the expected relation between stellar mass and SFR (the main sequence of star-forming galaxies, \citealt{Noeske2007}). We highlight that while the IllustrisTNG model has been calibrated to reproduce several galaxy scaling relations (e.g. the stellar mass-halo mass relation, \citealt{Pillepich2018}), the stellar mass-SFR relation was not invoked. On the other hand, the two free parameters of the radiative transfer postprocessing (the dust-to-metal ratio $f_\mathrm{dust}$ and the clearing timescale for the birth clouds of star-forming regions $\tau$) have been calibrated to reproduce various flux and color relations from the DustPedia sample in \citet{Trcka2022}, including a WISE W1-FUV relation which is very similar to the one presented here (see Section~\ref{sec:VISTA K vs. WISE W1} for our reasoning why to replace WISE W1 with VISTA $K$ as stellar mass tracer).

\begin{figure*}
    \centering
    \includegraphics[width=0.97\textwidth]{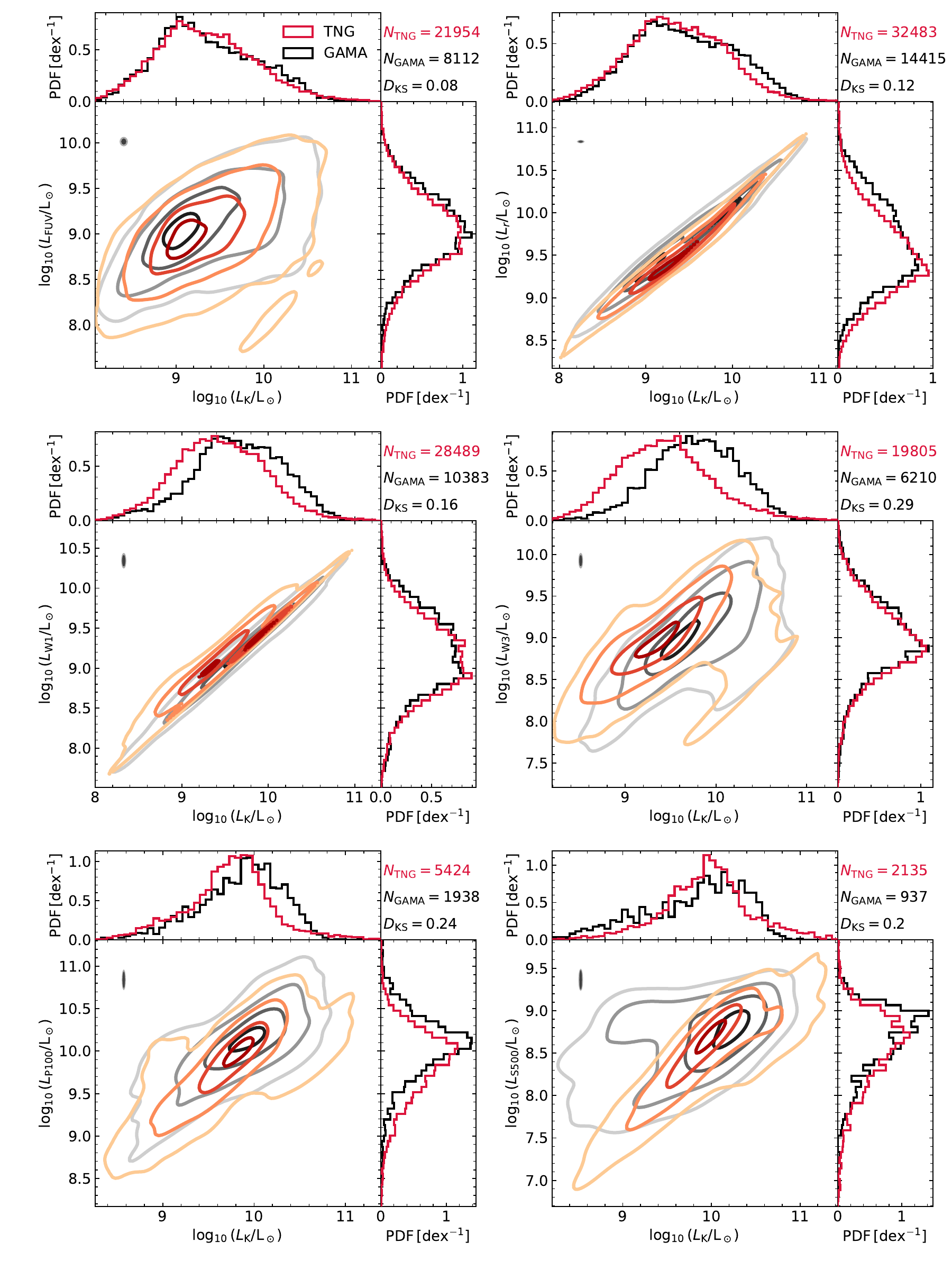}
    \caption{Six different flux-flux relations, for TNG100 (red) and observational data from the GAMA survey (black), for $0.002\leq z\leq0.1$ and $M_\star\geq10^{8.5}\,\mathrm{M}_\odot$.}
    \label{fig:fluxes}
\end{figure*}
\begin{figure*}
  \contcaption{The panels always have the VISTA $K$-band flux on the $x$-axis, and feature various bands (increasing with wavelength from the top left to the bottom right panel) on the $y$-axis. For both datasets, we filter out galaxies which lie below specific flux thresholds in any of the bands (see text for details). The number of remaining galaxies is given in the top right corner of each panel. The 2D distribution is estimated using a kernel density estimate (KDE). The different levels correspond to 5, 25, 60, and 90\,\% of the total KDE density. 1D color histograms for both datasets are also shown. Note that we use observer-frame fluxes here. An estimate of the average noise in the observations is indicated by the grey ellipses, with the darker (lighter) ellipse indicating the median (upper quartile) 1-$\sigma$ observational error bar. $D_\mathrm{KS}$ indicates the distance between the two distributions according to a two-dimensional Kolmogorov-Smirnov test. The flux-flux relations seen in the GAMA data are well reproduced by the TNG100 galaxies.}
\end{figure*}

\subsubsection{VISTA K vs. SDSS r}\label{sec:VISTA K vs. SDSS r}

The SDSS $r$-band luminosity also traces stellar mass (e.g. \citealt{Mahajan2018}), but due to increased dust attenuation and variability with stellar age it is less often used as a direct stellar mass proxy compared to the $K$ band (\citealt{Bell2003}). On the other hand, the stellar evolution templates in the NIR carry systematic uncertainties related to TP-AGB stars (\citealt{Maraston2006}; \citealt{Taylor2011}). We find that the $r$ and $K$-band fluxes correlate very tightly, in a similar fashion for both the GAMA and TNG100 data. The TNG100 galaxies are redder by $\approx0.25\,\mathrm{mag}$ which could be due to an overly effective dust attenuation in the $r$ band. Comparatively older or more metal-rich stellar populations in TNG100 could also contribute to this discrepancy, but \citet{TNG_Nelson} find that the TNG100 galaxy ages and stellar metallicities broadly agree with observational SDSS data within the systematic uncertainties (see their figure 2). Lastly, we find that systematic uncertainties of the SED templates for the evolved stellar populations are of the order of $\approx0.2\,\mathrm{mag}$ when testing different template libraries.

\subsubsection{VISTA K vs. WISE W1}\label{sec:VISTA K vs. WISE W1}

Since the WISE W1 flux traces the Rayleigh-Jeans tail of evolved stars, this band can also be used as a stellar mass estimate (e.g. \citealt{Jarrett2013}; \citealt{Meidt2014}; \citealt{Jarrett2023}; \citealt{Sureshkumar2023}). The comparison with GAMA fluxes reveals that there is a sizeable population of TNG100 galaxies above the GAMA distribution. While the 1D histograms indicate that this offset seems to be mostly due to the $K$-band flux being too low in TNG100, we caution that a strong selection effect is at play: only galaxies that have both $K$ and WISE W1 fluxes above the thresholds shown in Table~\ref{tab:FluxLimits} are included in the plot. If the selection is dominated by the WISE W1 band, and if the TNG100 galaxies are systematically brighter in this band than GAMA galaxies (at a fixed $K$-band luminosity), then this `WISE W1 excess' could manifest itself as a `VISTA $K$ deficiency' - even if the TNG100 and GAMA $K$-band distributions would match exactly. This is because TNG100 galaxies which are comparatively faint in the $K$-band can reach the required WISE W1 flux threshold, while GAMA galaxies at similar $K$-band luminosities would be discarded leading to the shown offset in the 1D $K$-band distributions. To visualize the flux-flux relations under this assumption of perfectly matching $K$-band luminosity distributions between TNG100 and GAMA, we show the results of a conditional KDE in Figure~\ref{fig:conditionalKDE}.

We find that the offset from the GAMA distribution strongly correlates with the number of star-forming regions (stellar populations with ages below 10 Myr which we model with the MAPPINGS-III templates) relative to the number of evolved stellar populations. At first sight, this suggests that the MAPPINGS-III templates are the cause of excess WISE W1 emission for the TNG100 galaxies. However, we found that the contribution of star-forming regions to the WISE W1 flux is small, typically below 5\,\% (see also Figure~\ref{fig:SEDbreakdown}). Instead, we suggest that emission from the diffuse dust causes the elevated WISE W1 fluxes, as the diffuse dust contribution also strongly correlates with the offset from the GAMA distribution and reaches values up to 70\,\%. Upon inspection of the simulated TNG100 spectra, we find emission features at the WISE W1 band which corresponds to the 3.3-micron polycyclic aromic hydrocarbon (PAH) feature (\citealt{Tokunaga1991}; \citealt{Kim2012}). It seems plausible that it is this PAH emission which causes the excess WISE W1 fluxes for the TNG100 galaxies, but whether this originates in overly emissive PAH dust in the THEMIS dust mix or if the MAPPINGS-III templates are overly effective in stochastically heating the surrounding diffuse dust remains unclear.

\subsubsection{VISTA K vs. WISE W3}\label{sec:VISTA K vs. WISE W3}

Since the WISE W3 band predominantly traces the PAH emission from PDRs (\citealt{Kapoor2023}), this flux is used as an alternative tracer for star-formation rate which is unaffected by dust attenuation (e.g. \citealt{Cluver2017}; \citealt{Elson2019}; \citealt{Naluminsa2021}; \citealt{Sureshkumar2023}). Similarly as in Section~\ref{sec:VISTA K vs. GALEX FUV}, we see the star-forming main sequence with similar slopes in the TNG100 and GAMA data, but with a clearer separation between the star-forming and quiescent galaxy populations compared to the $K$-FUV relation. The TNG100 galaxies populating the sequence in the bottom right corner, with WISE W3 luminosities 1.5\,dex below the main sequence, are all devoid of star-forming gas (i.e. have zero star-formation rate). This population of quiescent galaxies is also seen in the GAMA data. 

On the other hand, the star-forming TNG100 galaxies are slightly offset towards the top left corner. The 1D WISE W3 distributions match to great precision, but the TNG100 VISTA $K$ luminosities seem to be offset to lower values compared to the GAMA data. This is exactly the same effect as discussed in Section~\ref{sec:VISTA K vs. WISE W1} (a TNG100 excess in WISE W3 flux disguised as a deficiency in $K$-band flux due to selection effects), and we speculate that it also has the same origin (an excess of PAH emission from the diffuse dust component) as the diffuse dust emission contributes at least $\approx60\,\%$ of the WISE W3 flux for all star-forming galaxies (see also Figure~\ref{fig:SEDbreakdown}).

\subsubsection{VISTA K vs. PACS 100}\label{sec:VISTA K vs. PACS 100}

Since the FIR dust emission peak is usually encompassed by the 100 and 160\,$\mu\mathrm{m}$ bands (\citealt{Cortese2014}), the PACS 100 flux traces relatively warm dust. The correlation of this flux with the $K$-band exhibits a similar slope and scatter in the TNG100 and GAMA distributions. The TNG100 PACS 100 fluxes are systematically smaller than the GAMA fluxes, but the offset is very small ($\approx0.1\,\mathrm{dex}$). We note that for this and the next panel involving FIR fluxes, the galaxy samples shrink substantially (c.f. the GAMA and TNG100 base samples of 17\,932 and 61\,076 galaxies, respectively).

\subsubsection{VISTA K vs. SPIRE 500}\label{sec:VISTA K vs. SPIRE 500}

The SPIRE 500 band traces relatively cold dust, and can be used as a dust mass proxy since the dust budget in the ISM is dominated by cold dust ($T\lesssim25\,\mathrm{K}$, \citealt{Dunne2001}) and the SPIRE 500 flux is less affected by dust temperature variations than for instance the SPIRE 250 flux (\citealt{Galametz2012}). Hence, the correlation between $K$ and SPIRE 500 flux is a purely observational counterpart of the physical non-linear relation between stellar and cold dust mass (e.g. \citealt{Cortese2012}). 

While we find that the TNG100 and GAMA flux distributions broadly agree in this flux-flux relation, there is a sizable population of GAMA galaxies at low $K$-band luminosities ($L_\mathrm{K}\sim10^9\,\mathrm{L}_\odot$) with substantially elevated SPIRE 500 fluxes (by approximately one order of magnitude). When replacing the SPIRE 500 with the SPIRE 250 band we find a better agreement ($D_\mathrm{KS}=0.13$), moreover the 2D KDE contours do not show a population of elevated SPIRE 250 fluxes for the GAMA galaxies. The SPIRE 500 band is (unlike the SPIRE 250 band) susceptible to submillimeter (submm) excess. This excess flux could be due to very cold dust shielded from starlight or changes in the emission properties of the dust grains at submm wavelengths, but the exact origin remains unknown (\citealt{Kirkpatrick2013}; \citealt{Hermelo2016}). As the cold ($T\lesssim8000\,\mathrm{K}$) ISM is not modelled explicitly in the IllustrisTNG model but treated according to the two-phase model of \citet{Springel2003}, the lack of a cold ISM component could explain the absence of this galaxy population with elevated SPIRE 500 fluxes in TNG100.

However, the SPIRE 500 fluxes are also known to suffer more from source confusion (\citealt{Rigby2011}) and are less reliable than the SPIRE 250 fluxes. We tested a more stringent SNR criterion of five (instead of three), which mostly affects the SPIRE 500 band. We find that the population of GAMA galaxies with elevated SPIRE 500 fluxes vanishes almost completely in this case\footnote{The impact on any of the other results is minor when using a more stringent SNR criterion of $\mathrm{SNR}>5$}, with $D_\mathrm{KS}$ never changing by more than 0.07 for any of the relations in Figures~\ref{fig:fluxes} and~\ref{fig:colors}. Hence, this particular tension is not robust and could be due to observational uncertainties.

\subsection{Galaxy color-color relations}\label{sec:Color-color relations}

\begin{figure*}
    \centering
    \includegraphics[width=\textwidth]{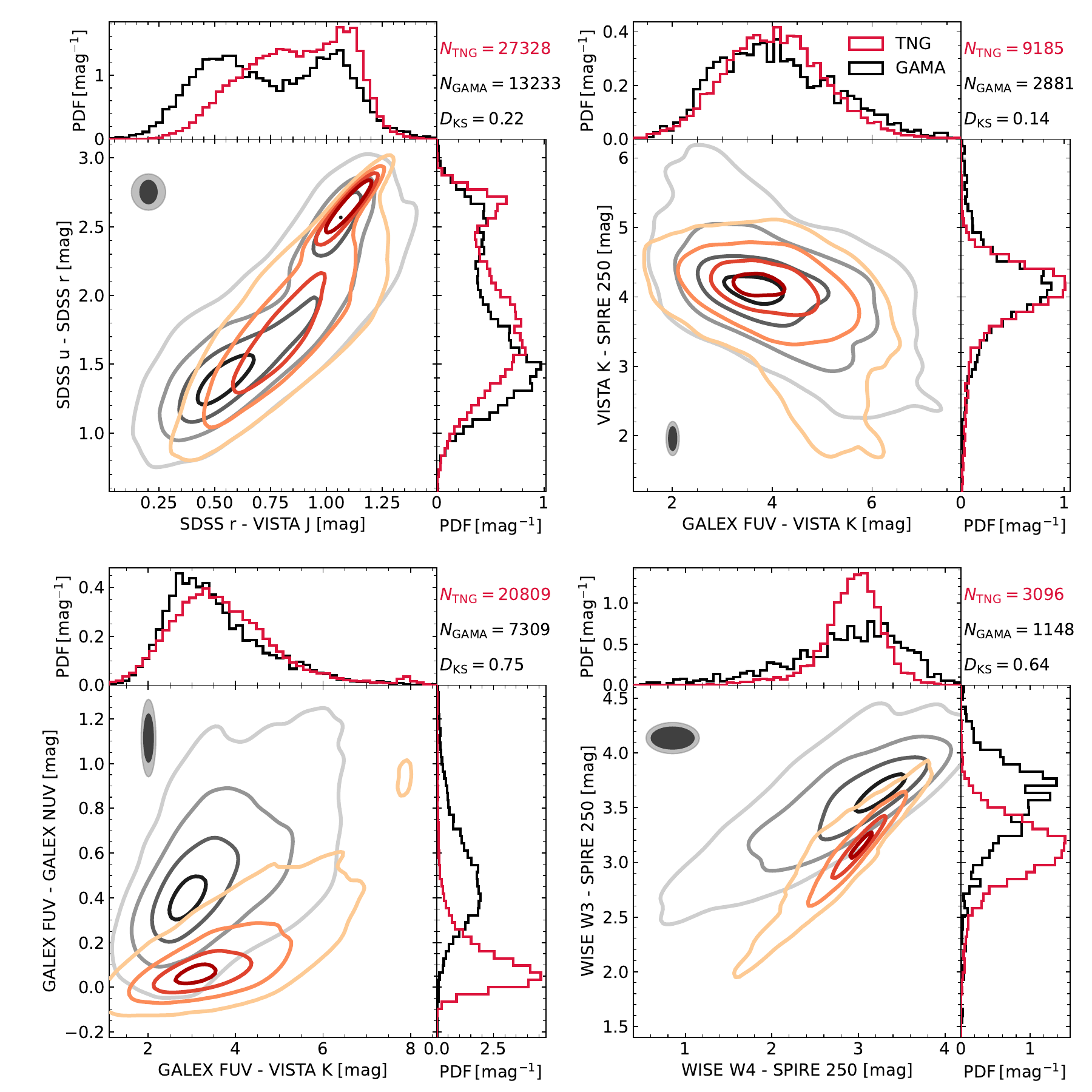}
    \caption{Four different color-color relations, for TNG100 (red) and observational data from the GAMA survey (black), for $0.002\leq z\leq0.1$ and $M_\star\geq10^{8.5}\,\mathrm{M}_\odot$. For both datasets, we filter out galaxies which lie below specific flux thresholds in any of the bands involved in the color-color relation (see text for details). The number of remaining galaxies is given in the top right corner of each panel. The 2D distribution is estimated using a kernel density estimate (KDE). The different levels correspond to 5, 25, 60, and 90\,\% of the total kernel density. 1D color histograms for both datasets are also shown. Note that we use observer-frame fluxes here. An estimate of the average noise in the observations is indicated by the grey ellipses, with the darker (lighter) ellipse indicating the median (upper quartile) 1-$\sigma$ observational error bar. $D_\mathrm{KS}$ indicates the distance between the two distributions according to a two-dimensional Kolmogorov-Smirnov test. TNG100 reproduces the observed color distributions in the two upper panels, but the TNG100 galaxies have flatter UV slopes and bluer WISE W3 - SPIRE 250 colors compared to the GAMA data.}
    \label{fig:colors}
\end{figure*}

We show four different color-color relations in Figure~\ref{fig:colors}. The galaxy samples are determined in the same way as in Figure~\ref{fig:fluxes}, i.e. they are derived from the GAMA and TNG100 base samples by imposing SNR and flux thresholds on each band involved in a specific color-color relation. An alternative realization of Figure~\ref{fig:colors} using TNG50 instead of TNG100 is shown in Figure~\ref{fig:ColorsTNG50}.

\subsubsection{(SDSS r - VISTA J) vs. (SDSS u - SDSS r)}

This color-color relation emulates the commonly used UVJ diagram (using the $V$-$J$ and $U$-$V$ Johnson filters), which is relevant due to its capability of separating the star-forming and quiescent galaxy populations observationally (\citealt{Williams2009}; \citealt{Whitaker2010}; \citealt{Patel2012}; see \citealt{Leja2019} for some limitations of the UVJ diagram). While dust attenuation shifts galaxies in the top right direction of the UVJ diagram, quiescent galaxies appear as a distinct population that is offset towards the top left direction. The UVJ diagram has also been studied in postprocessed cosmological simulations for TNG100 (\citealt{Donnari2019}; \citealt{Nagaraj2022}), TNG50 (\citealt{Baes2024}),  and SIMBA (\citealt{Akins2022}). Using a raytracing postprocessing method developed by \citet{TNG_Nelson}, \citet{Donnari2019} derive the rest-frame UVJ diagram for TNG100 at $z=0$ and find that it is broadly consistent with observational data, but do not compare the simulated and observed color distributions in detail.

In our case, we find two galaxy populations that are clearly separated as seen in observations. As we know the star-formation rates for the TNG100 galaxies, we can verify if these two populations indeed correspond to star-forming and quiescent galaxies. When splitting the galaxy population by specific star-formation rate (sSFR), we find that star-forming galaxies (with $\mathrm{sSFR}>10^{-10.5}\,\mathrm{yr}^{-1}$) indeed occupy the peak at blue colors and broadly extend to the top right corner, while quiescent galaxies (with $\mathrm{sSFR}<10^{-11.5}\,\mathrm{yr}^{-1}$ are located along a very narrow sequence offset from the star-forming sequence to redder $u$-$r$ colors.

However, the star-forming sequence appears to be slightly too red in TNG100 by $\approx0.25\,\mathrm{mag}$ along both axes. Multiple effects could contribute to render the star-forming galaxies too red for TNG100 (as discussed in section~\ref{sec:VISTA K vs. SDSS r}): at these wavelengths, the amount of dust as well as the dust model affects the colors. Furthermore, the SED template libraries for the evolved stars can affect the UVJ colors by up to 0.2\,mag (G. Worthey, private communication). And lastly, the stellar populations of the star-forming TNG100 galaxies could also be intrinsically too old or metal-rich. However, we remark that the $u$-$r$ color of $z=0$ TNG100 galaxies postprocessed with the simpler method of \citet{TNG_Nelson} reproduce observational data from SDSS, i.e. the star-forming galaxies are bluer by $\approx0.25\,\mathrm{mag}$ compared to our TNG100 colors calculated with SKIRT. At the same time, the quiescent galaxies which are less affected by dust attenuation are slightly redder (by $\approx0.1\,\mathrm{mag}$) in \citet{TNG_Nelson} compared to our results. This points towards too much dust reddening in our SKIRT pipeline, which is puzzling given the excellent agreement of our SKIRT fluxes with other flux-flux/color-color relations and luminosity functions. We defer a more detailed assessment of the intrinsic and dust-reddened optical colors of TNG100 galaxies to future work.

\subsubsection{(GALEX FUV - VISTA K) vs. (VISTA K - SPIRE 250)}\label{sec:Color relation b}

As discussed in Sections~\ref{sec:VISTA K vs. GALEX FUV} and~\ref{sec:VISTA K vs. SPIRE 500}, the GALEX FUV and SPIRE 250 fluxes trace galaxy SFR and dust mass, respectively. Hence, this color-color relation is an analogue of the physical galaxy scaling relation between specific star-formation rate and specific dust mass (e.g. \citealt{Cortese2012}; \citealt{Remy-Ruyer2015}; \citealt{Nanni2020}; \citealt{Shivaei2022}). Both GAMA and TNG100 feature a mild correlation between these colors, with a tail extending towards the bottom left corner which contains quiescent galaxies with very small specific dust masses. The peaks, widths, and correlation of the GAMA and TNG100 color distributions match to great precision for this relation.

We briefly compare this result to a similar color-color relation which was used to calibrate the SKIRT postprocessing parameters in \citet{Trcka2022} (their figure 6, panel g). Their color-color relation slightly differs from the one shown here as \citet{Trcka2022} adopted the WISE W1 band to trace stellar mass, while we use the VISTA $K$ band. As discussed in Section~\ref{sec:VISTA K vs. WISE W1}, the WISE W1 flux can contain significant PAH contribution from the diffuse dust component. We also examined the exact same color-color relation replacing the VISTA $K$ with the WISE W1 band, to reproduce the color-color relation that was used by \citet{Trcka2022} in the calibration process. We find for our datasets that the excellent match between the GAMA and TNG100 distributions vanishes ($D_\mathrm{KS}=0.6$), with the WISE W1-SPIRE 250 (GALEX FUV-WISE W1) color of GAMA being significantly redder bluer by $\approx0.5\,\mathrm{mag}$ compared to TNG100. This means that the radiative transfer calibration of \citet{Trcka2022} (using DustPedia data and WISE W1 fluxes) produces sensible results when using GAMA data and VISTA $K$ fluxes as shown here, but is in tension when using the same GAMA data with WISE W1 fluxes.

We found two different effects with coincidentally similar magnitudes, which can explain these discrepant results: first, the TNG100 WISE W1 fluxes contain PAH emission that seems to be too strong in the SKIRT setup used here (see Section~\ref{sec:VISTA K vs. WISE W1}). Second, the DustPedia WISE W1-SPIRE 250 colors are bluer by $\approx0.5\,\mathrm{mag}$ compared to the GAMA colors, probably related to selection effects (the DustPedia archive is a much smaller sample of 814 local galaxies with Herschel and WISE W1 detections). These two effects conspire to give consistent results for this color-color relation using the WISE W1 band and DustPedia or using the VISTA $K$ band and GAMA data.

\subsubsection{(GALEX FUV - VISTA K) vs. (GALEX FUV - GALEX NUV)}

This color-color relation has the UV slope GALEX FUV-NUV on the $y$-axis. Since the UV is dominated by star-forming regions and dust attenuation is very strong at these wavelengths, the UV slope of the TNG100 galaxies sensitively depends on the treatment of star-forming regions and the subsequent attenuation in the diffuse ISM. Hence, the UV slope $\beta$ is correlated with the infrared excess IRX (ratio of IR and UV luminosity) and commonly used as a measure for attenuation in the ISM using the IRX-$\beta$ relation (\citealt{Calzetti1997}; \citealt{Meurer1999}). We examine the FUV-NUV color as a function of FUV-VISTA $K$, which we use as a proxy for $\mathrm{sSFR}^{-1}$ as in Section~\ref{sec:Color relation b}.

We find that the sSFR and UV slopes are anticorrelated in both datasets, but the anticorrelation is substantially stronger in the GAMA data. Furthermore, the TNG100 UV slopes are also offset to lower values, with the peaks of the distributions differing by $\approx0.4\,\mathrm{mag}$. We also note that the FUV-NUV distribution of the GAMA galaxies is wider, which is (at least partially) caused by the relatively high noise levels of this particular color. 

When calculating FUV-NUV colors without diffuse dust component for the TNG100 galaxies we find that the FUV-NUV colors hardly change, meaning that the diffuse dust has a negligible impact on the UV slope. Instead, the FUV-NUV color is driven by the SED templates of both the evolved stellar populations and the star-forming regions, which contribute roughly similar fractions to the total UV fluxes (see Figure~\ref{fig:SEDbreakdown}). A redder FUV-NUV color (i.e. a steeper UV slope) could for instance be obtained with a more selective extinction in the FUV band from the dusty birth clouds in the MAPPINGS-III templates. Kapoor et al. (in prep.) find that for the 30 MW-like galaxies from the AURIGA simulation (\citealt{Grand2017}), the recent TODDLERS library for star-forming regions (\citealt{Kapoor2023}) yields redder FUV-NUV colors of $\approx0.15\,\mathrm{mag}$ compared to MAPPINGS-III. Whether this change fully resolves the tension in this color-color relation or additional adjustments need to be made (e.g. in the templates for the evolved stellar populations, which can also contribute substantially to the UV fluxes) would require postprocessing the TNG100 galaxies again varying the SED templates of the star-forming regions and evolved stellar populations, which is beyond the scope of this study.

\subsubsection{(WISE W4 - SPIRE 250) vs. (WISE W3 - SPIRE 250)}

Lastly, we show a color-color relation involving the SPIRE 250, WISE W3 and WISE W4 fluxes. As discussed in Section~\ref{sec:VISTA K vs. WISE W3}, the WISE W3 band traces PAH emission from the diffuse dust component. The WISE W4 flux originates from hot dust around star-forming regions (\citealt{Kapoor2023}), and we find that it comes roughly in equal parts from the MAPPINGS-III star-forming regions and the diffuse dust (see Figure~\ref{fig:SEDbreakdown}). Hence, this color-color relation measures the amount of hot dust and PAH emission relative to cold dust traced by SPIRE 250.

This relation is observationally particularly challenging to measure, resulting in the large observational errors and wider GAMA distributions. While the WISE W4-SPIRE 250 color distributions broadly match, the TNG100 WISE W3-SPIRE 250 colors are bluer by $\approx0.5$\,mag which we attribute to elevated WISE W3 fluxes due to PAH emission from the diffuse dust (as discussed in Section~\ref{sec:VISTA K vs. WISE W3}). The 2D distributions show that the slope of the relation is steeper in TNG100. This is expected since galaxies with high WISE W4-SPIRE 250 colors have a comparatively large fraction of their dust heated to high temperatures due to emission from star-forming regions. This in turn leads to a stronger WISE W3 excess for those galaxies and thus a steepening of this color-color relation for TNG100 galaxies.

\section{Summary}\label{sec:Conclusions}

We applied the radiative transfer postprocessing method developed by \citet{Trcka2022}, where the TNG50 simulation was analyzed, to the fiducial TNG100 run of the IllustrisTNG suite. The postprocessing method uses the dust MCRT code SKIRT to propagate the emission from evolved stars and star-forming regions through the dusty ISM. We generated broadband fluxes and low-resolution SEDs from the UV to the far-IR for all TNG100 galaxies in the $z=0$ and $z=0.1$ snapshots resolved by more than $\approx200$ star particles ($M_\star>10^{8.5}\,\mathrm{M}_\odot$), leading to a sample of $\approx60\,000$ postprocessed galaxies. This dataset (as well as the TNG50 and TNG50-2 fluxes and SEDs generated by \citealt{Trcka2022}) is publicly available on the IllustrisTNG website\footnote{\url{www.tng-project.org/gebek24}}. To test the fidelity of the cosmological simulation and our postprocessing method, we compared the simulated fluxes to low-redshift observational data. The following points summarize our main findings:

\begin{itemize}
    \item TNG100 luminosity functions from the UV to the far-IR fall within the range of low-redshift observational results (Figure~\ref{fig:LFs}). Residual discrepancies at the bright end in the UV/optical/NIR are on the level of systematic effects in the observations related to aperture choices. As noted by \citet{Trcka2022}, the improvement over the TNG50 simulation stems from the fact that the IllustrisTNG model was designed at the resolution of TNG100, i.e. the subgrid parameters were chosen such that TNG100 reproduces some key statistics of the low-redshift galaxy population (e.g. the stellar mass-halo mass relation).

    \item We compare six different flux-flux relations between TNG100 and observational data from GAMA in Figure~\ref{fig:fluxes}. To mimic the strong observational selection effects, we redistribute the TNG100 galaxies to arbitrary redshifts to compute a realistic apparent brightness distribution (Section~\ref{sec:Sensitivity}). Exploring the fluxes in various bands as a function of $K$-band luminosity (which traces stellar mass), we find a broad baseline agreement between TNG100 and GAMA. Tension in the WISE bands is correlated with the abundance of star-forming regions in TNG100 galaxies and with emission from the diffuse dust component. Hence, we attribute this tension to excess PAH emission, potentially related to overly effective stochastic dust heating from the star-forming regions. 

    \item Lastly, we use the same method applied for the flux-flux relations to compare four different color-color relations between TNG100 and GAMA. Tension exists mostly in the UV slope (TNG100 galaxies exhibit flatter UV slopes, i.e. lower FUV-NUV colors, than GAMA data) and in IR colors involving WISE bands. The former could be related to the extinction in the dusty birth clouds of the MAPPINGS-III templates not being selective enough, while the latter is again caused by excess PAH emission from the diffuse dust. However, we remark that uncertainties in the dust model, dust distribution, and templates for evolved stellar populations could also play a role.

\end{itemize}

We conclude that this low-redshift dataset provides a useful resource to test the fidelity of TNG100, explore observational systematics (e.g. aperture, inclination, or sample selection effects), and interpret the complexity faced in the observed galaxy population. Fundamentally, this is made possible by shifting the simulated data into the `observational realm'. This approach is complementary to studies in the `physical realm', and we highlight the importance of considering both approaches as they carry different systematics and biases. The dataset presented in this study represents an important step towards analyzing the vast IllustrisTNG simulation landscape in the `observational realm'.

\section*{Acknowledgements}

We thank Eric Rohr and Peter Camps for enlightening discussions. We also wish to express our gratitude towards the anonymous referee, whose feedback substantially improved the quality of this paper.

AG gratefully acknowledges financial support from the Fund for Scientific Research Flanders (FWO-Vlaanderen, project FWO.3F0.2021.0030.01).

This study made extensive use of the \texttt{Python} programming language, especially the \texttt{numpy} (\citealt{numpy}), \texttt{matplotlib} (\citealt{matplotlib}), and \texttt{scipy} (\citealt{scipy}) packages. We also acknowledge the use of the Topcat visualization tool (\citealt{Taylor2005}) and the \texttt{ndtest} \texttt{Python} package (\url{https://github.com/syrte/ndtest}).

The IllustrisTNG simulations were undertaken with compute time awarded by the Gauss Centre for Supercomputing (GCS) under GCS Large-Scale Projects GCS-ILLU and GCS-DWAR on the GCS share of the supercomputer Hazel Hen at the High Performance Computing Center Stuttgart (HLRS), as well as on the machines of the Max Planck Computing and Data Facility (MPCDF) in Garching, Germany.

GAMA is a joint European-Australasian project based around a spectroscopic campaign using the Anglo-Australian Telescope. The GAMA input catalogue is based on data taken from the Sloan Digital Sky Survey and the UKIRT Infrared Deep Sky Survey. Complementary imaging of the GAMA regions is being obtained by a number of independent survey programmes including GALEX MIS, VST KiDS, VISTA VIKING, WISE, Herschel-ATLAS, GMRT and ASKAP providing UV to radio coverage. GAMA is funded by the STFC (UK), the ARC (Australia), the AAO, and the participating institutions. The GAMA website is \url{http://www.gama-survey.org/}. We also use VISTA VIKING data from the GAMA database, based on observations made with ESO Telescopes at the La Silla Paranal Observatory under programme ID 179.A-2004.

This research has made use of the Spanish Virtual Observatory (\url{https://svo.cab.inta-csic.es}, \citealp{Rodrigo2012, Rodrigo2020}) project funded by MCIN/AEI/10.13039/501100011033/ through grant PID2020-112949GB-I00.

\section*{Data Availability}

The IllustrisTNG data used in this work as well as the generated broadband fluxes are publicly available at \url{https://www.tng-project.org/} as described by \citet{Nelson2019a}.

The GAMA data is publicly available as part of data release 4 (DR4, \citealt{Driver2022}) of the GAMA survey. DR4 can be accessed at \url{http://www.gama-survey.org/dr4/}.

All other data (observational luminosity functions, derived data for GAMA) and the analysis scripts are publicly available at \url{https://github.com/andreagebek/TNG100colors}.



\bibliographystyle{mnras}
\bibliography{main} 




\appendix

\section{Resolution effects: Comparison to TNG50}\label{app:resolution}

\begin{figure*}
    \centering
    \includegraphics[width=0.97\textwidth]{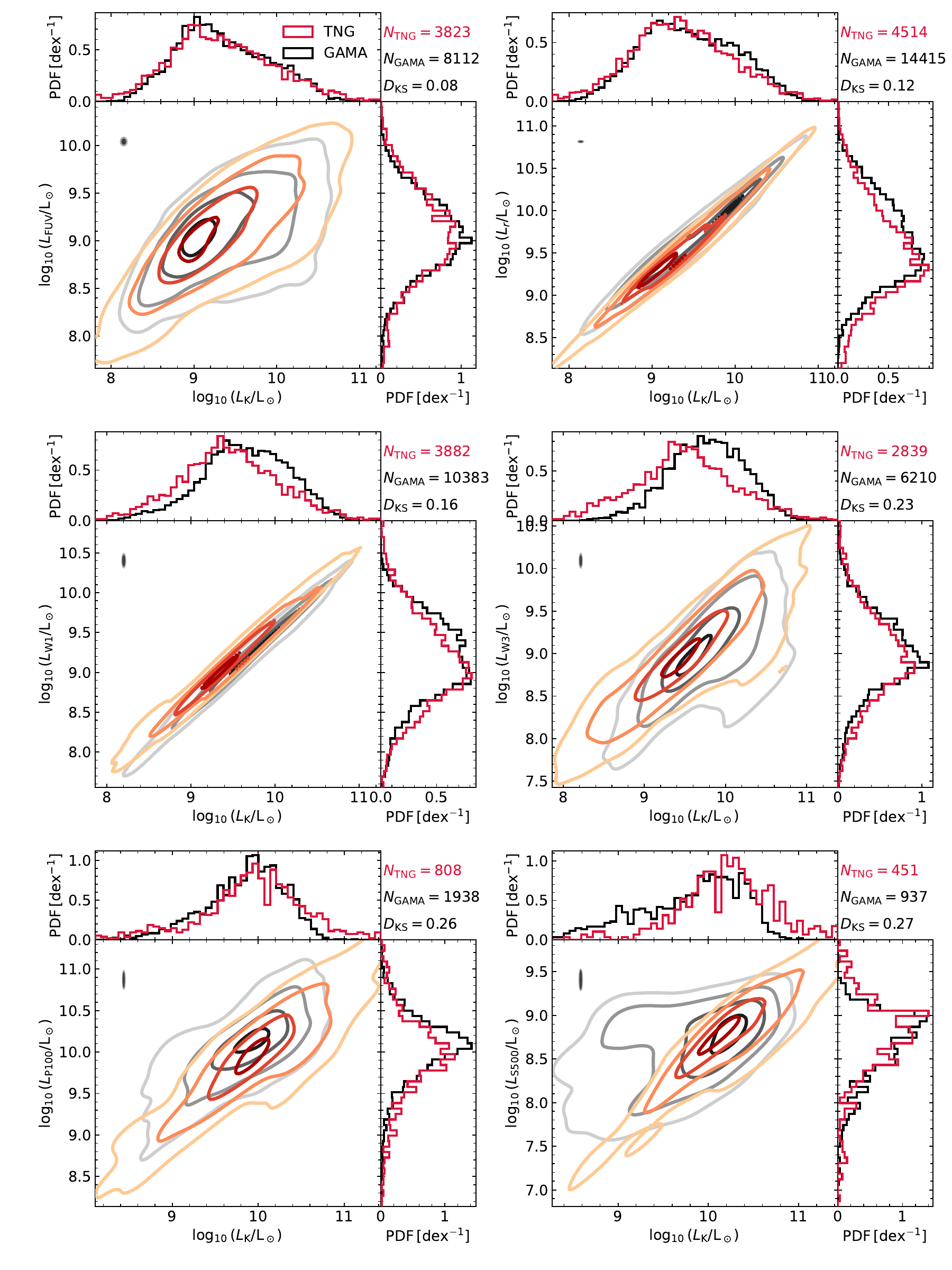}
    \caption{Same as Figure~\ref{fig:fluxes}, but using TNG50 instead of TNG100.}
    \label{fig:FluxesTNG50}
\end{figure*}

\begin{figure*}
    \centering
    \includegraphics[width=\textwidth]{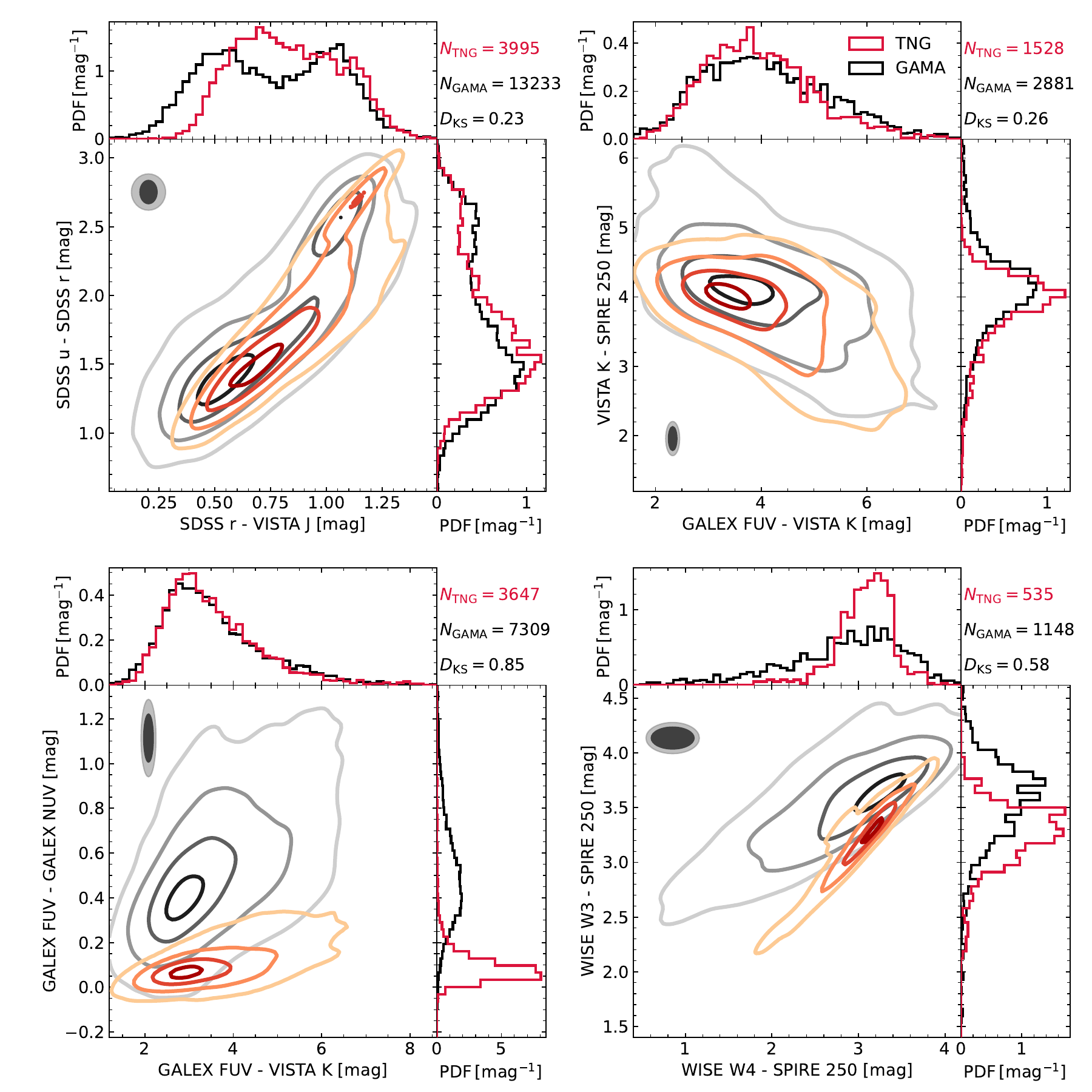}
    \caption{Same as Figure~\ref{fig:colors}, but using TNG50 instead of TNG100.}
    \label{fig:ColorsTNG50}
\end{figure*}

As shown in Figure~\ref{fig:LFs}, the IllustrisTNG fluxes exhibit a dependency on the simulation resolution. We show the flux-flux and color-color relations for TNG50 in Figures~\ref{fig:FluxesTNG50} and~\ref{fig:ColorsTNG50} to assess how the simulation resolution impacts our results. For the flux-flux relation, we find that the results up to the WISE W1 band are unchanged. The TNG50 WISE W3 (SPIRE 500) fluxes are slightly lower (higher) at fixed $K$-band luminosity compared to TNG100. For the color-color relations, the differences are more pronounced. TNG50 exhibits bluer VISTA $K$-SPIRE 250 colors, lower UV slopes, and bluer WISE W3-SPIRE 250 colors. Still, the level of agreement with GAMA for these flux-flux and color-color relations is comparable for TNG100 and TNG50, unlike the luminosity functions where TNG100 provides a significantly better match to the observational data. This indicates that to first order, the higher resolution of TNG50 leads to the galaxies being brighter at all wavelengths at fixed dark matter halo mass, while changes in the colors are only a higher-order (though non-negligible) resolution effect.

\section{Conditional flux-flux relations}

\begin{figure*}
    \centering
    \includegraphics[width=0.97\textwidth]{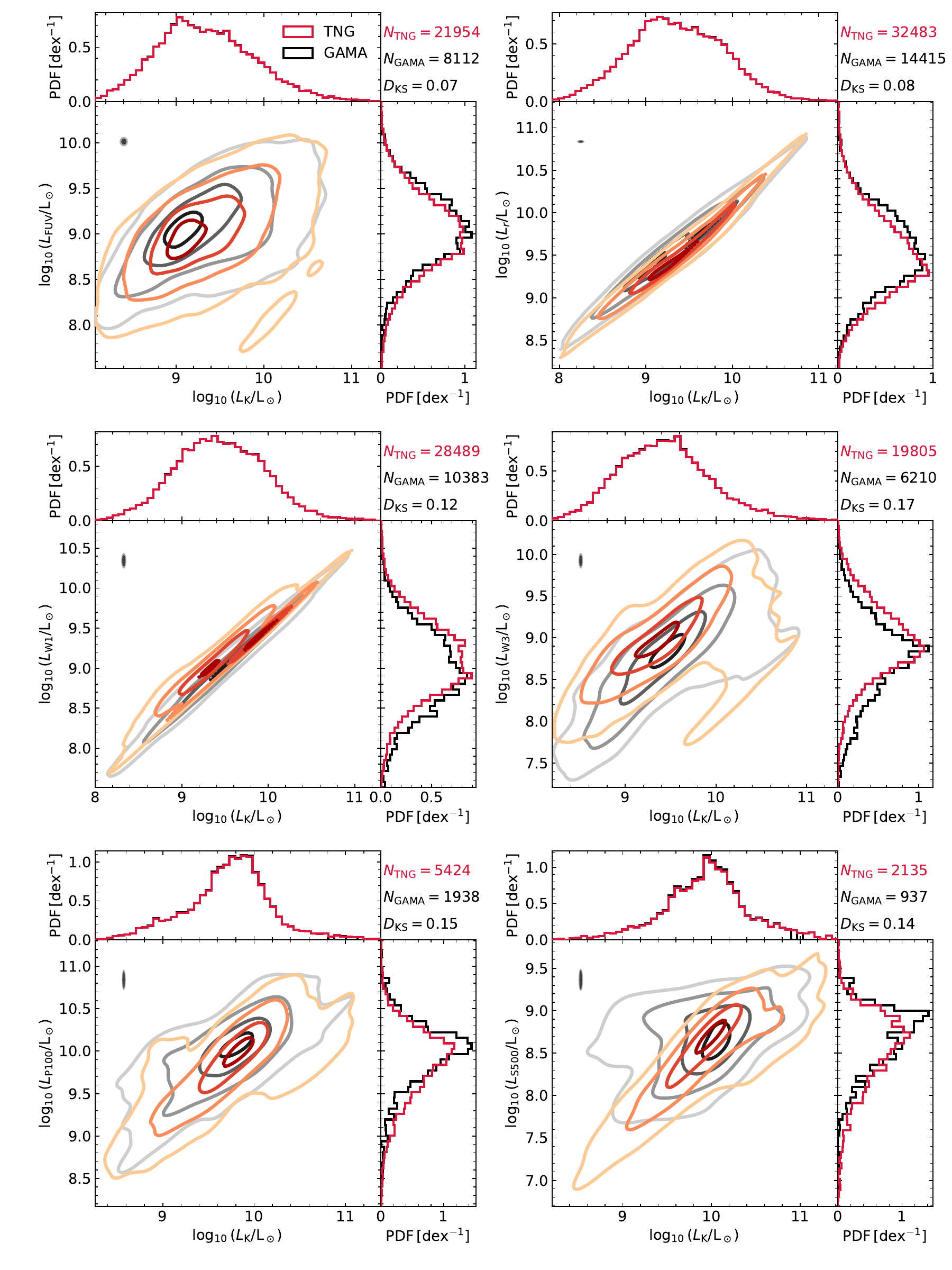}
    \caption{Same as Figure~\ref{fig:fluxes}, but using weights for the GAMA galaxies such that the VISTA $K$ distribution exactly matches the TNG100 result.}
    \label{fig:conditionalKDE}
\end{figure*}

As discussed in Section~\ref{sec:VISTA K vs. WISE W1}, imposing flux thresholds to select TNG100 galaxies potentially leads to biased $K$-band distributions if the TNG100 and GAMA colors are different. To visualize this effect and explore the differences in various bands under the assumption that the $K$-band distributions of TNG100 and GAMA galaxies perfectly match, we show the flux-flux relations with conditional KDEs in Figure~\ref{fig:conditionalKDE}. In this figure, we weigh all GAMA galaxies such that they reproduce the 1D $K$-band distributions of TNG100. The $D_\mathrm{KS}$ values correspond to the weighted one-dimensional test here, as opposed to the unweighted two-dimensional test used for all other flux-flux and color-color results. Comparing this result to Figure~\ref{fig:fluxes}, we note some differences especially for the center panels involving WISE bands. We refer to Section~\ref{sec:VISTA K vs. WISE W1} for a discussion on the cause and implications of this effect.

\bsp	
\label{lastpage}
\end{document}